\documentclass{ieeeaccess}
\usepackage{graphicx}
\usepackage{textcomp}
\def\BibTeX{{\rm B\kern-.05em{\sc i\kern-.025em b}\kern-.08em
    T\kern-.1667em\lower.7ex\hbox{E}\kern-.125emX}}

\usepackage{algorithm}
\usepackage{algpseudocode}
\usepackage{hyperref}
\usepackage{times}
\usepackage{amsfonts}
\usepackage{amsmath}
\usepackage{fancyvrb}
\usepackage{epsfig}
\usepackage{enumerate}
\usepackage{xcolor,colortbl}
\usepackage{amssymb}
\usepackage{latexsym,color,amscd,textcomp,url}
\usepackage{rotating}
\usepackage{slashbox}

\usepackage{algorithm}
\usepackage{algpseudocode}
\usepackage{epsfig}
\usepackage{subfigure}
\usepackage{calc}
\usepackage{amsmath}
\usepackage{multirow}
\usepackage{adjustbox,lipsum}

\newcounter{countdefinitions}
\newcounter{counttheorems}
\newcommand\revtext[1]{{\color{black}#1}}
\newcommand\srevtext[1]{{\color{black}#1}}

\begin{document}

\title{
PRESTvO: PRivacy Enabled  Smartphone based access To vehicle On-board units
 }

\history{Date of publication 19 June, 2020, date of current version 19 June, 2020.}
\doi{10.1109/ACCESS.2020.3003574}

\author{\uppercase{Bogdan Groza}\authorrefmark{1},\uppercase{Tudor Andreica}\authorrefmark{1},
\uppercase{Adriana Berdich\authorrefmark{1}, Pal-Stefan Murvay\authorrefmark{1} and Horatiu Gurban}.\authorrefmark{1}}
\address[1]{Faculty of Automatics and Computers, Politehnica University of Timisoara, 300223 Timisoara, Romania, e-mail: \{bogdan.groza,  tudor.andreica, adriana.berdich, pal-stefan.murvay, eugen.gurban\}@aut.upt.ro}
\tfootnote{
This work was supported by a grant of Ministery of Research and Innovation, CNCS-UEFISCDI, project number PN-III-P1-1.1-TE-2016-1317, within PNCDI III and by a grant of the Romanian Ministery of Research and Innovation, project number 10PFE/16.10.2018, PERFORM-TECH-UPT - The increasing of the institutional performance of the Polytechnic University of Timișoara by strengthening the research, development and technological transfer capacity in the field of "Energy, Environment and Climate Change", within Program 1 - Development of the national system of Research and Development, Subprogram 1.2 - Institutional Performance - Institutional Development Projects - Excellence Funding Projects in RDI, PNCDI III. 
}

\markboth
{Groza \headeretal: PRESTvO: PRivacy Enabled  Smartphone based access To vehicle On-board units}
{Groza \headeretal: PRESTvO: PRivacy Enabled  Smartphone based access To vehicle On-board units}

\corresp{Corresponding author: Bogdan Groza (e-mail: bogdan.groza@aut.upt.ro).}

\begin{abstract}
Smartphones are quickly moving toward complementing or even replacing traditional car keys. 
We advocate a role-based access control policy mixed with attributes that facilitates access to various functionalities of vehicular on-board units from smartphones. We use a rights-based access control policy for in-vehicle functionalities similar to the case of a file allocation table of a contemporary OS, in which read, write or execute operations can be performed over various vehicle functions. 
Further, to assure the appropriate security, we develop a protocol suite using identity-based cryptography and we rely on group signatures which preserve the anonymity of group members thus assuring privacy and traceability. To prove the feasibility of our approach, we develop a proof-of-concept implementation with modern smartphones, aftermarket Android head-units and test computational feasibility on a real-world in-vehicle controller. Our implementation relies on state-of-the-art cryptography, including traditional building blocks and more modern pairing-friendly curves, which facilitate the adoption of group signatures and identity-based cryptography in automotive-based scenarios.
\end{abstract}


\titlepgskip=-15pt

\maketitle

\section{Introduction and motivation}

The generous interface of modern smartphones and their ubiquitousness opens road for adding access control to various car functionalities as well as for remote configuration and rights delegation. 
In contrast, classical radio-frequency (RF) and/or mechanical vehicle keys are rigid and lack in terms of flexibility and functionalities. 
Perhaps surprising, despite their simplicity, classical RF keys have shown numerous flaws that led to a plethora of reported attacks
\srevtext{targeting weaknesses in regular RF keys \cite{Verdult12}, \cite{Wetzels14},
open-source immobilizer specifications \cite{Tillich12}, 
or passive keyless entry systems \cite{Francillon11}, \cite{Garcia16}, \cite{Wouters2019}. 
So it seems that the security of traditional car keys is lacking in many respects.}
The causes are numerous, including poor selection of cryptographic algorithms or poor randomness, etc. This merely complements a landscape which became familiar to us in the recent years as cars are unsatisfactory prepared in terms of security, e.g., \cite{Koscher10}, \cite{Checkoway11}, \cite{Miller14}.

By using smartphones, specific applications can be targeted and the interface customized to gain access to virtually any device or component from the car. Moreover, rights delegation can address complex scenarios due to increased connectivity at a global scale.
\srevtext{Consequently,} replacing traditional keys with smartphones appears like a natural step in achieving increased usability and an improved user experience. This is in fact proved both by many research works (which we separately address in the related work section) but also by recent industry efforts such as the Car Connectivity Consortium which drives a global initiative of top players from the automotive domain for car-to-smartphone connectivity\footnote{https://carconnectivity.org/}.  
\srevtext{To place the current research into context, Figure \ref{fig:interface} provides} a depiction of the interface that we implemented in PRESTvO. Some car functionalities are outlined and four user roles are displayed: car owner, driver, passenger and a technician role. Table \ref{tab:role_rights} summarizes role rights which are marked in a similar fashion to access rights over files in a modern operating system.

\newcommand\rr{$\mathsf{r}$}
\newcommand\ww{$\mathsf{w}$}
\newcommand\ee{$\mathsf{e}$}
\newcommand\nn{$\mathsf{-}$}

\Figure[t!](topskip=0pt, botskip=0pt, midskip=0pt)[width=6cm]{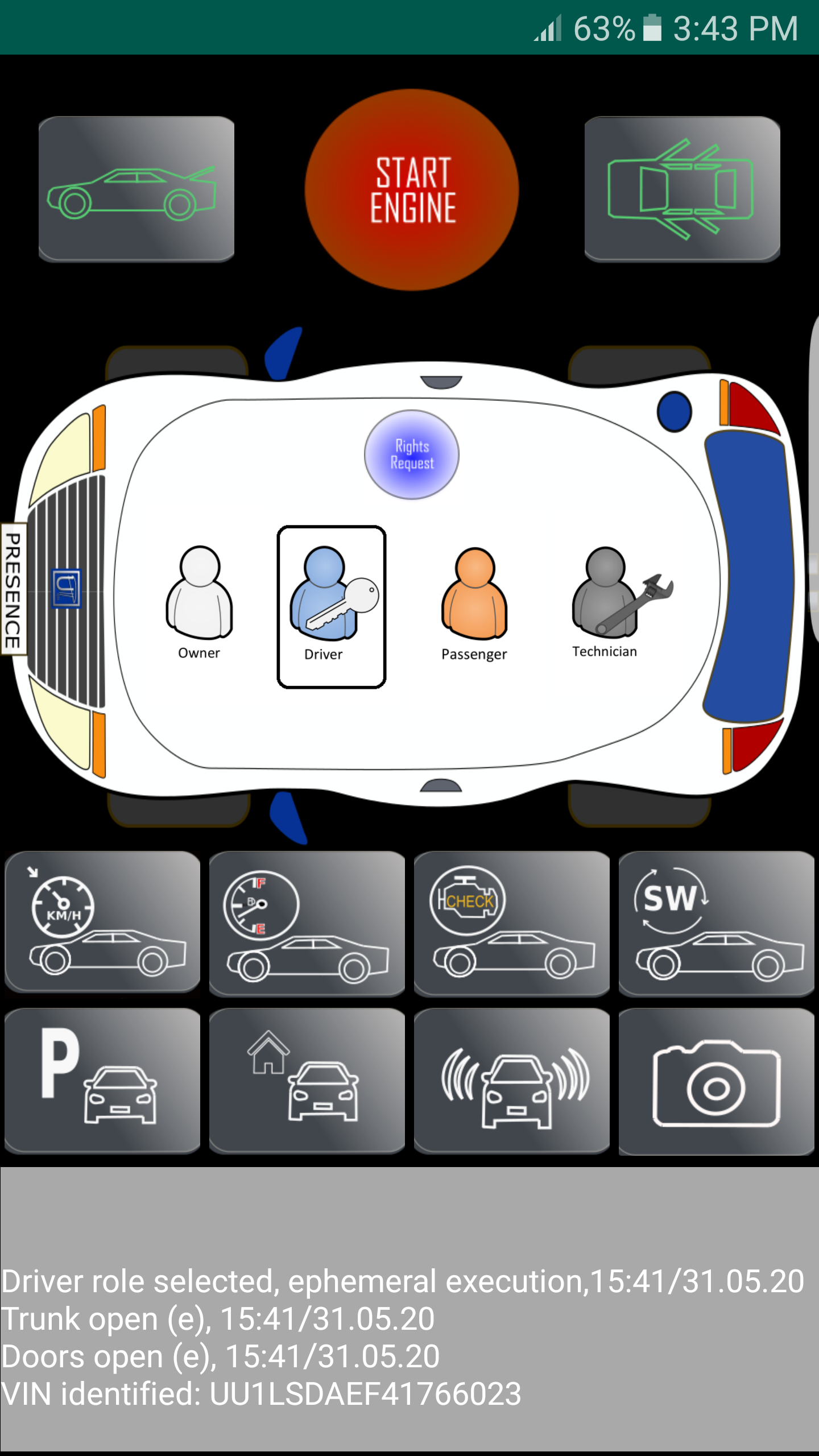}{PRESTvO user interface.\label{fig:interface}}

\begin{table}[tb!]
\scriptsize
	\centering
    \begin{center}
	\caption{An example of role access rights}
	\label{tab:role_rights}
		 \begin{tabular}
		 {  >{\arraybackslash}p{1.7cm} c r r r r r  } 
		\hline \hline
		\textbf{ } & \rotatebox{90}{\textbf{Owner}}& \rotatebox{90}{\textbf{Driver}} & \rotatebox{90}{\textbf{Technician}}  & \rotatebox{90}{\textbf{Child Ocuppant}} & \rotatebox{90}{\textbf{Valet}} & \rotatebox{90}{\textbf{Passenger}}  \\[0.5ex]	
		\hline  \hline	

 \rowcolor[gray]{.95}

Start Engine 	& \nn\nn\nn 	& \nn\nn\ee		& \nn\nn\ee 	& \nn\nn\nn  	& \nn\nn\ee 	& \nn\nn\nn  \\
Open Trunk 		& \nn\nn\ee		& \nn\nn\ee		& \nn\nn\ee	 	& \nn\nn\nn 	& \nn\nn\nn 	& \nn\nn\ee	 \\

\rowcolor[gray]{.95}
Open Doors 		& \nn\nn\ee		& \nn\nn\ee		& \nn\nn\ee	 	& \nn\nn\ee		& \nn\nn\ee		& \nn\nn\ee	\\

\rowcolor[gray]{.95}
Limit speed 	& \rr\ww\nn  	& \rr\ww\nn  	& \nn\nn\nn  	& \nn\nn\nn 	& \nn\nn\nn 	& \nn\nn\nn \\

Fuel Level  	& \rr\nn\nn 	& \rr\nn\nn 	& \rr\nn\nn 	& \nn\nn\nn 	& \rr\nn\nn 	& \nn\nn\nn \\

Diagnosis		& \nn\nn\ee 	& \nn\nn\ee 	& \nn\nn\ee 	& \nn\nn\nn 	& \nn\nn\nn 	& \nn\nn\nn \\

SW Update 		& \nn\nn\nn 	& \nn\nn\nn 	& \nn\nn\ee 	& \nn\nn\nn 	& \nn\nn\nn 	& \nn\nn\nn \\

\rowcolor[gray]{.95}

Park car 		& \nn\nn\nn 	& \nn\nn\ee 	& \nn\nn\ee 	& \nn\nn\nn 	& \nn\nn\ee 	& \nn\nn\nn \\

\rowcolor[gray]{.95}
Home  			& \nn\nn\ee 	& \nn\nn\ee 	& \nn\nn\ee 	& \nn\nn\nn 	& \nn\nn\nn 	& \nn\nn\nn \\

Alarm 			& \nn\nn\ee 	& \nn\nn\ee 	& \nn\nn\ee 	& \nn\nn\nn 	& \nn\nn\ee 	& \nn\nn\nn \\

Start A/C 		& \nn\nn\ee 	& \nn\nn\ee 	& \nn\nn\ee 	& \nn\nn\nn 	& \nn\nn\ee 	& \nn\nn\ee \\

\rowcolor[gray]{.95}
Defrost  		& \nn\nn\ee 	& \nn\nn\ee 	& \nn\nn\ee 	& \nn\nn\nn 	& \nn\nn\ee 	& \nn\nn\nn \\

\rowcolor[gray]{.95}
Mirrors 		& \nn\nn\ee 	& \nn\nn\ee 	& \nn\nn\ee 	& \nn\nn\nn 	& \nn\nn\ee 	& \nn\nn\nn \\
Lights			& \nn\nn\ee 	& \nn\nn\ee 	& \nn\nn\ee 	& \nn\nn\nn 	& \nn\nn\ee 	& \nn\nn\nn \\

\rowcolor[gray]{.95}
Play music 		& \nn\nn\ee		& \nn\nn\ee		& \nn\nn\ee	 	& \nn\nn\ee		& \nn\nn\ee		& \nn\nn\ee	\\

Limit volume	& \rr\ww\nn  	& \rr\ww\nn  	& \rr\ww\nn  	& \nn\nn\nn 	& \nn\nn\nn 	& \nn\nn\nn \\

Trip Computer  	& \rr\ww\nn 	& \rr\ww\nn 	& \rr\nn\nn 	& \nn\nn\nn 	& \nn\nn\nn 	& \nn\nn\nn \\

			\hline
			\hline
		\end{tabular}
	\end{center}
\end{table}

However, replacing traditional keys with smartphones comes with additional security and privacy challenges. \srevtext{For example, smartphones will have to pair with the car over a wireless communication interface such as Wifi, Bluetooth or NFC. But all these interfaces have been commonly found vulnerable, e.g., key reinstallation attacks on the WPA2 have attracted much attention a few years ago \cite{Vanhoef17}, some vulnerabilities of NFC-based payments were shown by \cite{Giese19}, and quite long list of potential exploits on Bluetooth can be found in \cite{Albahar16}, \cite{Hassan18} with more recent results showing attacks on Bluetooth elliptical curve based pairings in \cite{Biham19}. 
Such vulnerabilities can be overcome only at the application layer by proper protocol designs based on specific cryptographic functionalities. This is precisely our intention here in PRESTvO, to design, implement and test a secure protocol for gaining access from a smartphone to a car via an existing wireless interface, e.g., WiFi or Bluetooth.
The protection mechanism will not be limited to the communication channel, it will also have to address on-device adversaries, such as malicious manufacturers, that may want to retrieve information from the vehicle in order to track the users. Recent incidents showed that major companies can be involved in privacy leakages either by their own consent or due to data breaches \cite{Isaak18}, \cite{Zou18}. Consequently, we focus both on the security of the communication channel and on the privacy of the users which is assured by the more demanding cryptographic group signatures. In this way, as the identity remains hidden behind the group key, the car should be unable to log any specific information about the person which gained access to the car. 
In what follows we briefly discuss the cryptographic toolset that enables us to reach the desired security and privacy goals.

}

\srevtext{\textit{Cryptographic building blocks}.   In principle, cryptography offers a comprehensive set of functions, i.e., cryptographic primitives,} that can assure various security objectives \srevtext{, e.g., authenticity, confidentiality, etc.} 
Besides regular symmetric and asymmetric primitives, i.e., \srevtext{public-key encryptions and digital signatures}, more recent advances set room for more exotic cryptographic functionalities. These include identity-based cryptography where the identity of a user can be used to derive his public key or group signatures where the identity of a user can be preserved anonymously under the public-key of a group (still allowing \srevtext{the group manager} to trace the user if a dispute arises).  
There is no question that these cryptographic functionalities will be sooner or later adopted by the industry. The AUTOSAR standard already defines interfaces for regular cryptographic building blocks in the automotive domain \cite{CAL}, \cite{CSM}. Identity-based signatures are part of an ISO standard \cite{ISOSig} published long ago and it targets embedded devices such as smartcards. There are numerous research works that advocate for the use of these cryptographic building blocks for car access-control, these are summarized in Table \ref{tab:related} and will be discussed in the related work section. 
However, the heterogeneity of the addressed environment, where smartphones interact with in-vehicle units, raises performance concerns. Our work builds upon various cryptographic building blocks (identity-based signatures, group signatures, etc.) and besides designing a car access protocol we try to bring answers regarding the feasibility of deploying this solution both on mobile devices and on in-vehicle components.

\begin{figure}[thb!]
\centering
\includegraphics[width=7 cm]{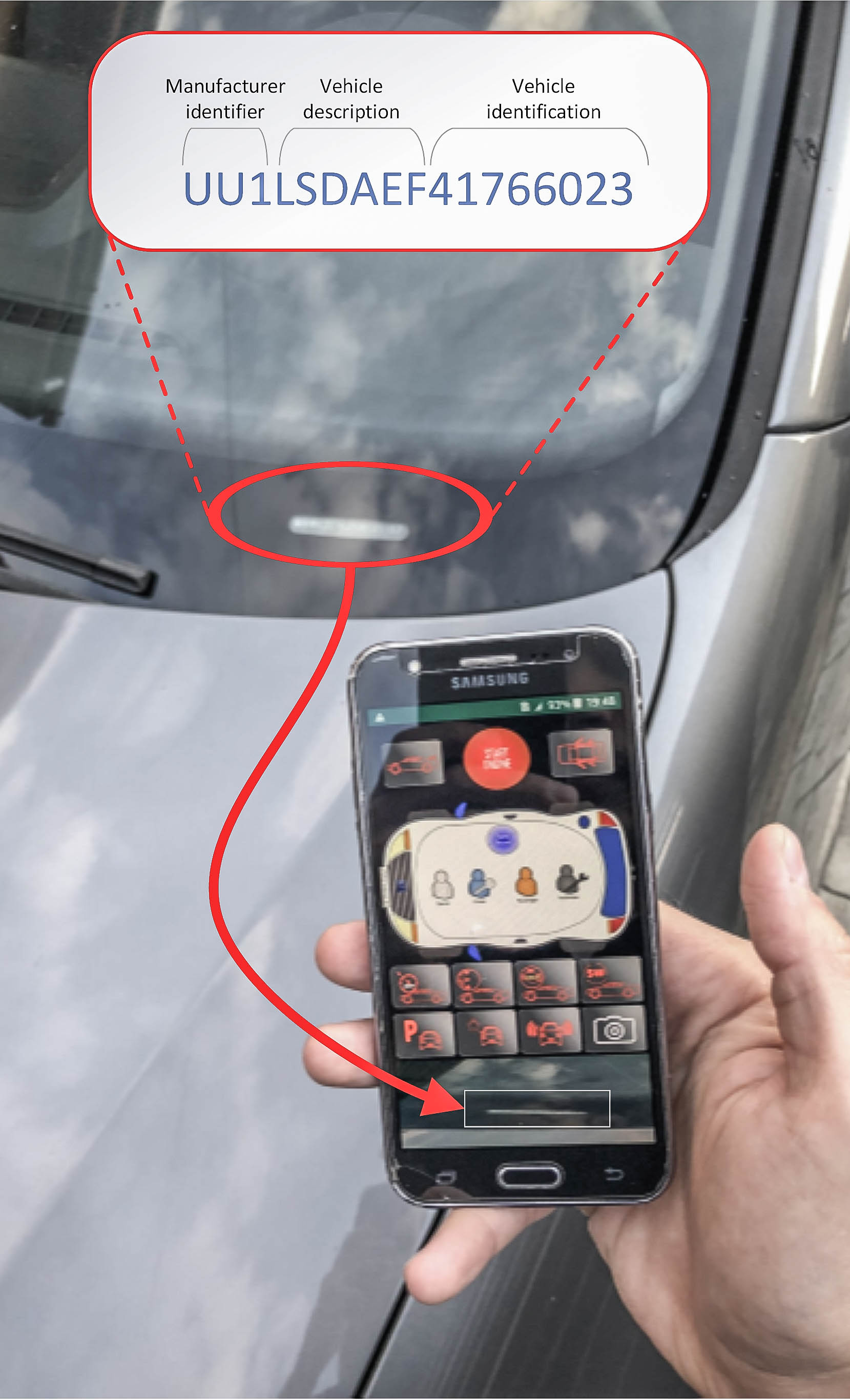}
\caption{User handling the PRESTvO application to colect the car VIN number, i.e., the identity of the car}
\label{fig:presto_vin}
\end{figure}

\srevtext{\textit{Garnering advantages from group and identity-based signatures.} While some of the cryptographic building blocks that we use are more demanding, e.g., group signatures, there are clear benefits behind using them in this car access scenario. By using group signatures, the car will be aware which role is accessing the car, i.e., owner, driver, passenger or technician, but will have no information on the entity that instantiated the specific role. That is, there may be multiple drivers, passenger, technicians and even car ownership may be shared, while the car (and implicitly the car manufacturer) will be unable to log information regarding the exact user (it is only the role which stays visible). With specific functionalities of group signatures, which are later discussed in the protocol design section, the exact user can still be traced by the group manager in case when a dispute arises. Our design emphasizes on the right of ownership and thus we let the car owner to be in possession of the group manager secret key (other deployments may choose to attribute this functionality to a trusted authority). 
While traditional signatures may preserve the anonymity of the users by using pseudonyms in the certificates, it is still possible to separate between users based on their distinct public keys and additional information, e.g., driving time and location leakages, may be corroborated for the de-anonymization of the user behind the pseudonym. Group signatures provide better privacy guarantees in this respect.

The use of identity-based signatures will make public-keys far easier and more intuitively to manage. For example, we show in Figure \ref{fig:presto_vin} a user attempting to collect the identity of a car from its VIN number that is located on the driver's side where the windshield meets the dashboard (this is not the only location but it is the most common for cars). 
To recognize the text from the VIN number, we used the text detection package from the Mobile Vision API\footnote{https://developers.google.com/vision} which provides a framework for objects identification in photos and video. The text of the VIN number from the dashboard was collected with almost 100\% accuracy showing that there are no technological shortcomings in this respect (provided that the dashboard is clean and the phone camera properly pointed to the VIN). 
Clearly, other elements, such as the license plate, can be used for the same purpose. The VIN however is immutable and will remain forever associated with a car, while the license plate can change. For example, in many countries new cars have temporary license plates until they are sold to the first customer. Thus the VIN number provides a more reliable identity for the car. This functionality may prove particularly useful in car-rental scenarios as well as for companies which share several cars between their employees. 
}

\begin{figure*}[th!]
\centering
\includegraphics[width=17 cm]{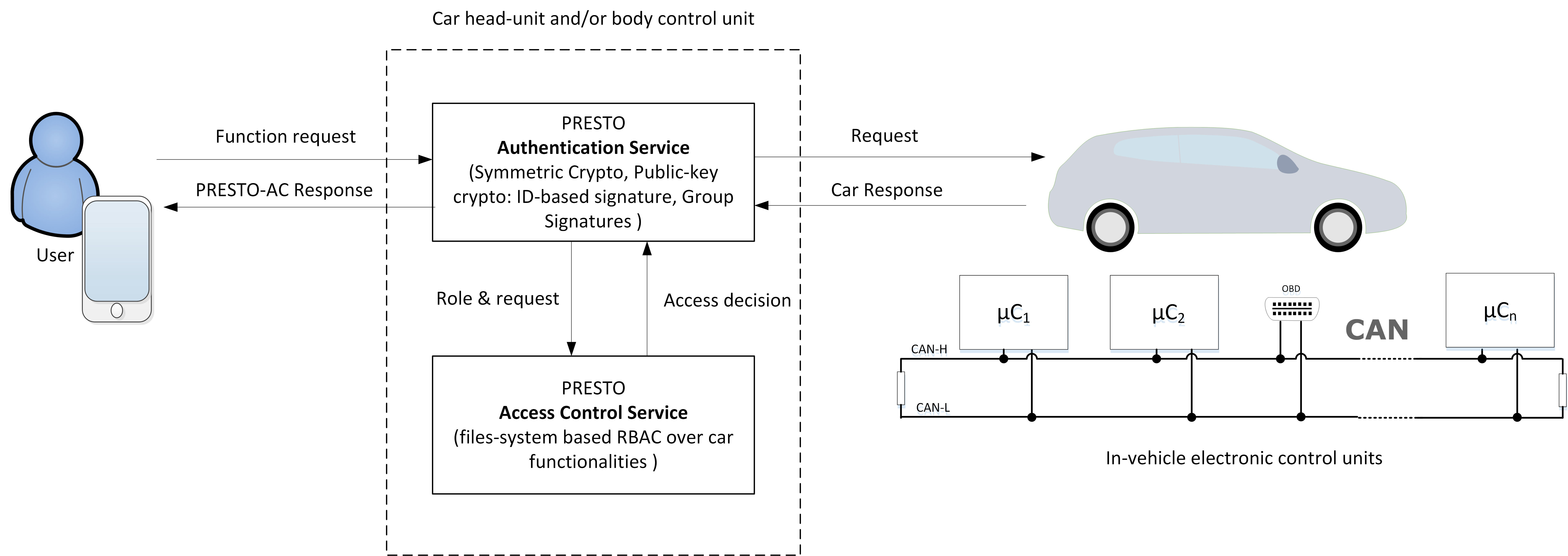}
\caption{PRESTvO system design}
\label{fig:system}
\end{figure*}

\textit{System design goals.} We now briefly discuss the design goals of our proposal. Figure \ref{fig:system} gives an overview of the addressed system. A user requests a particular functionality of the car which is to be executed by some in-vehicle electronic control unit (ECU). Access is mediated by PRESTvO. First an authentication service is called which verifies the identity of the user and the role he invoked. If the identity and roles are verified, the role along with the request is passed to the access control service which in turn verifies authorization for the particular request and returns the access decision. In case of a positive decision, the request is passed to the car which in turn responds according to the request, i.e., by executing the particular functionality and sending a response message. The user receives a response from PRESTvO which may be negative if his function request could not be approved or a confirmation otherwise.  We design the protocol behind PRESTvO with both security and privacy in mind and also without forgetting that we address functionalities inside a car and target real-world automotive-grade embedded devices. The following summarizes the goals of our work:

\begin{enumerate}

\item \emph{secure access control} to all vehicle functionalities mediated by the use of smartphones is the prime intention of our work,

\item a \emph{flexible access control policy} determined by roles and attributes falling middle of the road between RBAC and attribute-based access control (ABAC) which seems the best option due to the variety of car-usage scenarios,

\item \emph{rights delegation} and also rights revocation directly from the smartphone is a natural functionality,

\item \emph{user privacy}, by which we keep the identity of the user anonymous to the car and manufacturers, is enforced by the use of group signatures which hide the identity of a specific user inside a group,

\item \emph{user traceability} in case when a dispute arises is a mandatory procedure due to legal implications, e.g., the car may be involved in an accident and it becomes necessary to be able to trace a particular user,

\item \emph{flexible use of wireless interfaces WiFi, Bluetooth and NFC} according to existing support on the hardware that we used (smartphones and vehicle head-units),

\item \emph{comprehensive performance tests} on real-world automotive grade controllers are a must in order to prove that implementation is realistic with respect to  state-of-the-art in automotive on-board units.

\end{enumerate}

\subsection{Related Work}

\newcommand\yes{\checkmark}
\newcommand\no{--}

\begin{table*}[h!]
	\centering
    \begin{center}
    \vspace*{-10pt}
	\caption{Summary of some existing proposals for car keys with enhanced capabilities (in chronological order by year of publication)}
	\label{tab:related}
		 \begin{tabular}
		 {  >{\arraybackslash}p{0.5cm} >{\arraybackslash}p{5cm} >{\arraybackslash}p{5cm} >{\centering\arraybackslash}p{0.8cm}  >{\centering\arraybackslash}p{2.2cm}>{\centering\arraybackslash}p{0.6cm}>{\centering\arraybackslash}p{0.8cm}  } 
		\hline \hline
		\textbf{Paper} &	 \textbf{Main Concept}& \textbf{Platform for Key Deployment} & \textbf{Comm. intrf.}  & \textbf{Other cryptographic rq.} & \textbf{Car AC} & \textbf{Rights deleg.}  \\[0.5ex]	
		\hline  \hline

\cite{dmitrienko2012smarttokens} &
Access control rights delegation with trusted execution environment on Android  
& Nexus S (Android 2.3.3 patched with TrustDroid security extensions)  
& NFC & \no & \no & \yes \\[0.5ex]

\rowcolor[gray]{.95}
\cite{Busold13}&
Car access and rights delegation from an Android smartphone, security reinforced by smart-card	
& Samsung Galaxy S3, Arduino Uno (8-bit microcontroller for car immobilizer)	
& NFC	& \no 	& \yes	& \yes   \\[0.5ex]
 
\cite{timpner13}&
Secure vehicle-to-cloud communications based (uses OAuth \& HSM) 	
& \no
& NFC	& \no 	&	\yes & \no    \\[0.5ex]

\rowcolor[gray]{.95}
\cite{Kasper13}&
Rights Management with NFC Smartphones and Electronic ID Cards
& Blackberry Bold 990
& NFC	
& \no	& \no	& \no   \\[0.5ex]

\cite{han2014mvsec}&
 Pairing cars and smartphones by OOB (out-of-band) channels: light and sound (augmented by the Encrypted Key Exchange protocol,  Diffie-Hellman version)
&Motorola Droid 1, Android 2.2.3 (Froyo)
& BT 	& \no 	& \yes	& \no   \\[0.5ex]

\rowcolor[gray]{.95}
\cite{symeonidis2017sepcar}&
Car sharing functionality with secure multiparty computation
& Intel Core i7, 2.6 Ghz CPU, 8GBof RAM
&NFC, BT  & Secure multi-party computation	& \yes	& \yes   \\[0.5ex]

\cite{dmitrienko2017secure}&
Car sharing by short-range wireless communication (RFID, NFC, BLE) with two-factor authentication
& Samsung S4 (Android 5.0), Google Nexus 5 Android 5.1 (Lollipop),  + Giesecke \& Devrient Mobile Security Card (MSC) with DESFire applet,  contactless Mifare DESFire EV1 smartcard, Samsung Galaxy Gear SM-V700 smartwatch (Android 4.2)
& NFC, BLE  & \no  & \yes & \no   \\[0.5ex]

\rowcolor[gray]{.95}
\cite{Wei17}&
Car sharing in a hierarchical system between KGC (key generation center), owner/rental company, user	
& Android Nexus 5 
& NFC & Identity-based encryption and signatures	& \yes 	& \yes   \\[0.5ex]

\cite{Groza17}&
Rights delegation to a car from a low-cost embedded platform by using only symmetric crypto-primitives	
&MSP430 (16-bit microcontroller from Texas Instruments for key), Freescale S12 (as car immobilizer) 
& RF & \no 	& \yes	& \yes   \\[0.5ex]

\rowcolor[gray]{.95}
Our Work & Car access and rights delegation, preserving anonymity by using group signatures, eliminating PKI by using identity-based cryptography

& LG, Samsung J5, ERISIN \& PNI car head-units, Infienon TC297 (for car immobilizer)	
& NFC, BT, WiFi	& Identity-based signatures, group signatures 	&	\yes & \yes   \\[0.5ex]

			\hline
			\hline
		\end{tabular}
	\end{center}
	\vspace*{-10pt}
\end{table*}

Related work on vehicle access control and rights delegation is extensive. \srevtext{While only a limited amount of research papers have been focusing on traditional car immobilizers, e.g., \cite{Lemke05}, there is quite a large number of recent works that address the use of smartphones for accessing vehicle functionalities. In what follows, we give a brief overview of the existing proposals and summarize the most relevant of them in Table \ref{tab:related}}. We also point out in the table whether the work uses any enhanced cryptographic capabilities, e.g., identity-based signatures, group signature, etc., besides the regular symmetric/asymmetric primitives which are present in all of the works.  

The use of smartphones for access-control systems inside buildings and as replacement of traditional physical keys has been explored as early as the works of \cite {Bauer02Comparing}, \cite{Bauer07Lessons}.

In \cite{Busold13} a full platform for car access and rights delegation from an Android smartphone is presented. The security is reinforced by the use of a smart-card and the authors present both a proof-of-concept implementation and strong security arguments by model checking with ProVerif.
A hierarchical car sharing architecture is proposed in \cite{Wei17}. The authors consider only a simplified hierarchy with 3 levels: a key generation center, the owner or the rental company and the end-user. A proof of concept implementation is presented on an Android Nexus 5 smartphone, the protocol relies on identity based encryptions and signatures. 
The authors of \cite{dmitrienko2012smarttokens} propose a generic smartphone-based NFC access control system that allows access rights delegation. The access control system is based on a multi-level smartphone security architecture designed to provide trusted execution and storage environments.  Formal security analysis as well as a proof of concept implementation on a simplified system model are provided.
Several NFC-based use cases for the automotive environment are also described and implemented in \cite{steffen10}. 

Another secure access control system for car sharing is proposed in \cite{dmitrienko2017secure}. It employs two-factor authentication provided by an RFID token and a soft token to enable access to offline cars. The proposed instantiation uses a secure execution platform that can be implemented on devices such as smartcards or smartwatches. Another approach for car sharing is proposed in \cite{symeonidis2017sepcar}. The paper presents a decentralised protocol that provides both security and privacy allowing owners to share their cars. Protocol analysis and a proof-of concept implementation are also considered.

Other works that use smartphones for gaining access to vehicles are  \cite{alli12} and \cite{bose11}.
The Green Move project described in \cite{alli12} is a vehicle sharing system. In this project, the vehicles are equipped with Green e-box devices, which communicate with a smartphone via Bluetooth and with the cloud (Green Move Center) via HTTP. Using the smartphone application, the user retrieves the valid electronic key from the Green Move Center, which contains an encrypted ticket, the start time of reservation as well as all the information to identify the car. The encrypted ticket is used to lock/unlock the doors.
In \cite{bose11}, the Terminal Mode technology is described. This technology integrates the smartphone into the car head unit. In this scenario, the input and output functions are the responsibilities of the car head unit, while the smartphone acts as the application platform. New functionalities can be added to the car head unit by easily upgrading the smartphone.

Some works are focused on cost-efficient solutions. The implementation of a dedicated device for car-rights delegation on low-cost MSP430 microcontrollers from Texas Instruments is discussed in \cite{Groza17}.
A keyless car access system using RFID cards (e-driver licenses) is proposed in \cite{huang18}. In this scenario, each driver license is assigned to the driver's identity based on an RFID card using a serial number stored in the cloud database. If the serial number exists in the database, the driver can use the car and the owner knows who is driving the car based on information provided by a smartphone application.

Pairing mobile devices with cars has also been targeted. A secure pairing between mobile devices and vehicles based on out-of-band (OOB) channels is proposed in \cite{han2014mvsec}. The authors present several key agreement protocols using light and sound as OOB channels. Protocol analysis with AVISPA \cite{Armando05} as well as implementations are also provided.

Privacy concerns for smartphone applications in the automotive domain have been also addressed for pay-by-phone parking systems \cite{Garra07} or GPS tracking  \cite{Li18}.

\subsection{Selecting setup components}

Table \ref{tab:device_spec} provides a summary of the devices that we used in our setup. In what follows, we discuss these in detail.

\begin{table*}[h!]

	\centering
    \begin{center}
    
	\caption{Devices from our experiments: smartphones, head-units and in-vehicle control unit}
	
	\label{tab:device_spec}
		 \begin{tabular}
		 {  l c >{\arraybackslash}p{3.7cm}  >{\centering\arraybackslash}p{3cm}>{\centering\arraybackslash}p{3cm}>{\centering\arraybackslash}p{2cm}  } 
		\hline \hline
		\textbf{Device} &	 \textbf{Android}& \textbf{CPU} & \textbf{Memory}  & \textbf{WiFi} & \textbf{BT}  \\
		\hline  \hline	

\rowcolor[gray]{.95}
LG Optimus P700	& 4.0.3	& 1.0 GHz Cortex-A5	& 4 GB (2.4 GB user available)  512 MB RAM	& Wi-Fi 802.11 b/g/n, Wi-Fi Direct, hotspot, DLNA	& 3.0, A2DP \\

Samsung J5 & 5.1.1 & Quad-core 1.2 GHz Cortex-A53 & 8 GB, 1.5 GB RAM & Wi-Fi 802.11 b/g/n, Wi-Fi Direct, hotspot & 4.1, A2DP \\

\rowcolor[gray]{.95}
Samsung S5 & 6.0.1 & Quad-core 2.5 GHz Krait 450 & 16 GB, 2 GB RAM & Wi-Fi 802.11 a/b/g/n/ac, dual-band, Wi-Fi Direct, hotspot & 4.0, A2DP, EDR, LE, aptX \\
		
Samsung S7 & 6.0.1 & Octa-core (4x2.3 GHz Mongoose \& 4x1.6 GHz Cortex-A53) & 32 GB, 4 GB RAM & Wi-Fi 802.11 a/b/g/n/ac, dual-band, Wi-Fi Direct, hotspot & 4.2, A2DP, LE, aptX \\

\rowcolor[gray]{.95}
Head-unit PNI/ERISIN	& 7.1.1	& Quad-core 1.63 GHz Cortex A7 &	12  GB, 1/2 GB RAM	&	Wi-Fi 802.11 b/g/n,  hotspot, &	4.0, A2DP, BR/EDR  	\\

Infineon TC 		&N/A							&	 Triple Core 300MHz TriCore					& 8 MB Flash, 728KB RAM 	&  	N/A	&N/A\\

			\hline
			\hline
		\end{tabular}
	\end{center}
\end{table*}

\emph{Android head units and smartphones.} For our experimental setup we acquired two Android head-units with similar computational/communication capabilities.
The first of them from ERISIN was designed to replace SEAT and VW head units. The head unit provides a 9-inch capacitive display with 1024 x 600 resolution running the Android 7.1 Nougat OS.
The CPU it uses is an Allwinner Quad-Core T3 SoC (Quad-core Cortex A7, 1.63GHz and Mali400 MP2 GPU), 2GB RAM and 16GB internal memory. The storage can be increased by using microSD cards or USB connected memory devices. The T3 SoC integrates a high number of peripherals providing support for a multitude of standards: USB, SATA, UART, TWI, SPI, EMAC, GMAC, PS2.
The unit offers wireless Bluetooth and Wi-FI 802.11b/g/n connectivity. In addition it integrates: GPS, AM/FM radio tuner, RDS, DAB/DAB+ and CAN communication. Also this unit includes an USB connected rear view camera. Diagnostics information can be retrieved by using the included Bluetooth OBD2 Module.
The second aftermarket head unit, PNI A8020 HD, is a generic replacement head unit. This is a lower cost version but is essentially deployed on almost the same hardware. It comes with a 7-inch capacitive display and 1GB RAM but it uses the same Allwinner Quad-Core T3 SoC along with the same level of connectivity.

\emph{Vehicle on-board units.}
The on-board unit functionality can be built either as a stand-alone unit or as part of an ECU responsible for several other functionalities related to the body domain. We advocate for the latter since car access control is traditionally implemented as part of the body control module (BCM). The protocols implemented by the car access functionality are based on computationally intensive public key cryptography. Moreover, an embedded platform suitable to serve as the on-board unit as well as to implement other vehicle body functions should be able to perform all its designated functionalities in a timely manner. This calls for the use of a high performance automotive grade embedded platform capable of performing public key cryptography operations. 
We selected the TC297, an Infineon Aurix microcontroller, which can act in a real car as the BCM with on-board unit functionality. The multicore architecture of Aurix 32 bit microcontrollers is built to offer high performance. Covering automotive communication technologies such as CAN (and CAN-FD), FlexRay and Ethernet, the TC297 is suitable for a wide range of automotive applications. Additionally, the Aurix family introduces a hardware security module which provides random number generation, AES128 HW acceleration and a trusted execution environment for cryptographic algorithms. All these features make the TC297 a suitable candidate for the designated application.

\subsection{Selecting communication interfaces}

A short discussion on the three communication interfaces, i.e.,  Bluetooth, WiFi and NFC, that we use in our deployment now follows. While each of them can be used for any of the protocol components, pros and cons exist. Also, some restrictions may occur due to the unavailability of some of them in existing components. For example, the head units that we use are not equipped with NFC readers while their Bluetooth connectivity allows only for media streaming. Next, we give a brief overview of these interfaces and how they are used in our practical implementation.

\emph{Bluetooth} is a technology for wireless data transfer between devices for a short range with low power consumption. The maximum packet size that can be transferred on Bluetooth BR/EDR is 1021 bytes. In the last years, Bluetooth technology has been frequently used to communicate between the user and the car. The main use case is for the car infotainment system and in-vehicle wearables applications but also for car access control and maintenance tools. Bluetooth devices use profiles to specify the features that are supported and the type of data that can be transmitted/received.
The infotainment units used in our work have four Bluetooth Profiles:  Advanced Audio Distribution Profile (A2DP), Audio/Video Remote Control Profile (AVRCP), Hands-Free Profile (HFP), and  Headset Profile (HSP). These profiles can be used only for multimedia functionalities. 
The smartphones that we used on this work  have a richer set of  Bluetooth profiles:
Advanced Audio Distribution Profile (A2DP), Hands-Free Profile (HFP), Headset Profile (HSP), 
File Transfer Profile (FTP), Message Access Profile (MAP), Object Push Profile (OPP), and Phone Book Access Profile (PBAP).
To transfer packets of bytes between two devices, both devices have to support at least one of the Bluetooth Profiles that uses the OBEX protocol (Object Exchange). The Bluetooth profiles that use OBEX are FTP, OPP, and MAP.
Since our infotainment units do not have any Bluetooth profile with OBEX support, we used Bluetooth only between smartphones and relied on WiFi for communicating with the infotainment unit as we discuss next. However, Infotainment units with OBEX support for Bluetooth exist, so this is not a technical limitation for our protocol, it is just a small limitation of our setup.

\textit {Wireless networking (Wi-Fi)} is a technology for communication with high speed data transfer. In automotive, Wi-Fi technology is sometimes used by infotainment systems and is an essential component for the connecting cars in vehicle-to-vehicle communication (V2V), vehicle-to-infrastructure (V2I) or vehicle to pedestrian  communications, etc. A number of recent works have also focused on the use of WiFi for phone-to-phone communication inside vehicles, e.g., \cite{Sun16}.  Consequently, we consider that deploying part of our protocol over WiFi is realistic and will benefit from a higher data rate. In particular, we find that Wi-Fi is specifically suitable for protocol components that rely on the larger group signatures.

\textit{Near-field communication (NFC)} is a set of communication protocols which offers the possibility to establish a short-range communication between two electronic devices. NFC relies on RFID, having the operating frequency at 13.56 MHz and the bit rate between 106 kbit/s and 424 kbit/s. The communication range is up to 20 cm. A full NFC capable device can operate in one of the following three modes: card emulation, reader/writer and peer-to-peer. The first mode permits a smart NFC-enabled devices to act like smart cards, the second mode may be used for the reading and writing of NFC tags and the last mode, peer-to-peer, offers the possibility for two NFC-enabled smart devices to communicate in an ad-hoc manner. Each device that participates to the communication can be either the initiator or the target. A passive target can be powered by the RF field generated by the initiator. When compared to Bluetooth, which is also a short-range communication technology, NFC operates at a much shorter range and at slower speeds. On the other hand, the power consumption is an advantage of NFC as it consumes less power than Bluetooth. Another advantage of NFC is that it doesn't require pairing and, from the security point of view, a shorter communication range may preclude adversaries from intercepting the communication channel. Still, attacks have been reported on NFC as well, e.g., \cite{Verdult11}, \cite{Francis10}. We do use NFC for sharing access rights between smartphones but still rely on cryptographic building blocks to assure security.

\srevtext{
\textit{Brief discussion on connectivity.} The specific use-case that we address, i.e., car access mediated by smartphones, calls only for short distance connectivity. That is, the range of Bluetooth and WiFi is generally restricted to 10-100 meters \cite{Webb07} with possible extensions to around 200 meters for WiFi when the devices are outdoors. NFC targets different type of applications and is limited to a few dozen centimeters. This coverage is of course sufficient for a user that tries to gain access to the car. Other actions, such as blacklisting identities or certificate, i.e., certificate revocations, or removing certain user rights from the car, may be more efficiently performed remotely. In this case 4G/5G connectivity will be needed. This type of connectivity is within reach for modern vehicles and is in fact commonly required for vehicle-to-everything (V2X) communication. There is a large body of research works focusing on vehicular ad-hoc networks, routing and even security and privacy issues for such scenarios, e.g., \cite{Chen17}, \cite{Muhammad18}, \cite{Lai20}. Our work will focus on Bluetooth and WiFi communication which are more convenient for the scenario that we address.
}

\section{Design concept}

\begin{figure*}[th!]
\centering
\includegraphics[width=18 cm]{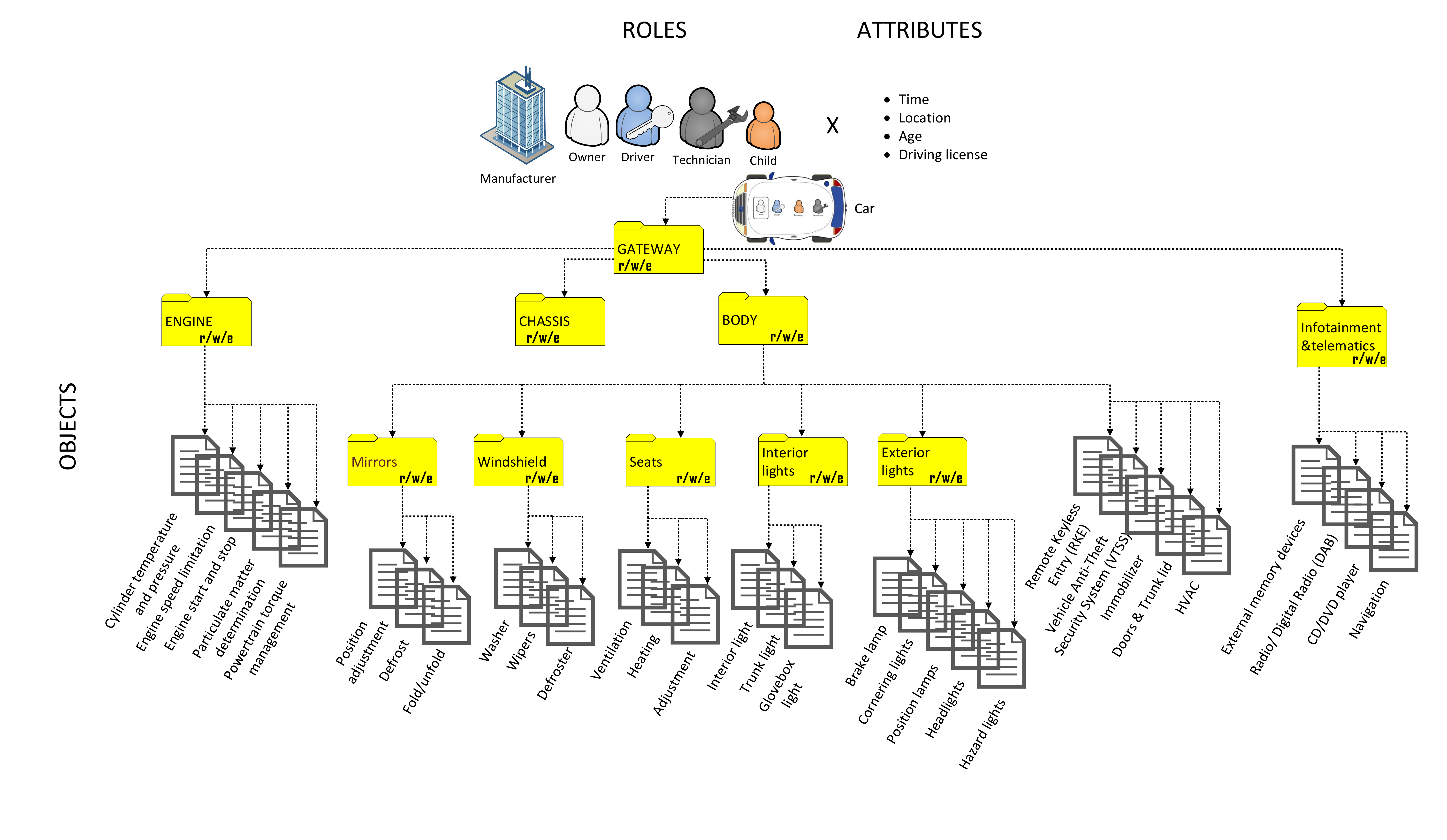}
\caption{Overview of role-based access control in PRESTvO}
\label{fig:presto_ac}
\end{figure*}

In this section we discuss the design concept behind our proposal. Subsequently, we give precise details on each component of the proposed protocol suite. 

\newcommand\perm{\mathsf{PERMISSIONS}}
\newcommand\object{\mathsf{OBJECTS}}
\newcommand\action{\mathsf{ACTIONS}}
\newcommand\awrite{\mathsf{w}}
\newcommand\aread{\mathsf{r}}
\newcommand\aexecute{\mathsf{e}}

\subsection{Access control concept}

We now present the concept behind our access control policy. The access control procedure is based on role-based access control (RBAC) to which we add some attributes that are needed for the roles. Using RBAC seems natural in automotive environments since manufacturers can easily define specific roles for a car, e.g., driver, passenger, child occupant, etc., and each role may offer specific access rights to users. \srevtext{Using Discretionary Access Control (DAC) or alternatively Mandatory Access Control (MAC) are also viable alternatives but associating rights to specific roles seems the more natural approach for our car access control scenario. Roles will also enforce the anonymity of each actor carrying the role under group signatures.} RBAC is well understood and standard specifications exist, e.g., \cite{Ferraiolo01}. Numerous extensions of it have also been discussed, e.g., using attributes \cite{Kuhn10}, location-aware policies \cite{Damiani07}, public-key certificates \cite{Chadwick03}, or trust \cite{Chakraborty06}, etc. This opens road for many future applications and while the basic RBAC may be somewhat rigid it can be easily augmented and made more flexible. 

A graphical depiction of the proposed access control model is suggested in Figure \ref{fig:presto_ac}. Following at least in part accepted/standardized definitions from \cite{Ferraiolo01}, we briefly summarize core elements:

\begin{enumerate}
\item \textit{Users} can either be individuals or entities instantiated by software agents, however, in the protocol descriptions that follow, we generally assume that users are persons requesting a particular action from the car,

\item \textit{Role} represents the role played by an entity (an individual or some software agent). Roles include car owners, drivers, technicians, child occupants, etc., which are all played by users. Other roles such as the car rental company or the manufacturer may be played by software agents that delegate rights over the car or execute various tasks, e.g., a software update, etc.,

\item \textit{Attributes} are characteristics associated to a user, they include: time, location, driver license, age, etc.  For example, a technician may perform a particular update or access to a component only if he is in the range of a particular location (e.g., the authorized garage). Attributes are either numerical or boolean, each attribute can be set to $\perp$ when a particular attribute is not applicable to a particular user or in a particular scenario. For example, it may be irrelevant whether a technician has or not a driving license or sometimes location information may be unavailable while certain rights should be executed on the car.

\item \textit{Objects} constitute the car functionalities intuitively viewed by us as files that may be classified into \textit{macro-objects} intuitively viewed by us as folders. Macro-objects are the car functional domains related to engine, chassis, body and infotainment. Objects are the associated functionalities, e.g., adjusts seats or lights, use the infotainment unit, etc. To simplify our model in Figure \ref{fig:presto_ac} we have only considered macro-objects associated to four functional domains: engine, chassis, body and infotainment. This instantiation may be easily extended, we present what it seems sufficient for most scenarios that we could imagine.

\item \textit{Actions} are the activities that can be performed on an object. We view permissions over car functionalities similar to traditional Unix-like systems. Similarly to Unix files and folders we consider three types of actions: read simply lists the content that is as available, write modifies specific values and execute is the ability to run the particular functionality. That is $\action=\{\awrite, \aread, \aexecute \}$ where each action is instantiated by a binary flag. 
The read permission allows a user read data, for example the user may read the fuel level, the mileage counter or the status of many other subsystems in the car. Write permissions are necessary to set specific values, for example setting the date and time, the cruise speed or resetting the trip computer. Execution rights, are required by specific programs, such as a movie player or by a software update. A technician may be entitled to make a software update, but not to play movies from Netflix, while for the passenger it's the reverse.

\item \textit{Permissions} are the authorization given to a role over an object which makes $\perm = 2^{\object \times \action}$.  For example, a potential car buyer may be authorized to list all functionalities in the car on his mobile phone, but without the possibility to run them. On the contrary, the manufacturer may be entitled to update the functionality. For simplicity, in the instantiation from  Figure \ref{fig:presto_ac} we consider that permissions from a macro-object propagate identically over the objects below, but this can be changed according to practical needs.

\end{enumerate}

\textit{Defining roles, persistent and ephemeral delegation.} We consider that roles are predefined by the producer during the manufacturing process. Subsequently, the car owner is responsible for assigning the group public keys for each role. In this way the manufacturer cannot control a car owned by some individual but it does have control over the rights given to each role which is necessary to avoid misuse of a particular service.  
We consider that the owner of the car assumes a root role and that the root is the only role allowed to install public keys in the car. \srevtext{Rights delegation can be persistent or ephemeral and it can be performed by any actor in a specific role. As we later present in the experimental results section, we allow delegation from one smart-phone to another via NFC, but of course any other interface such as WiFi, Bluetooth or even 4G can be used for this purpose. We prefer NFC due to its short range which makes it harder for an adversary to eavesdrop on the channel (the delegation protocol is secure nonetheless, so any communication channel can be used)}. Ephemeral delegation is designed to be short lived, e.g., a car that is rent for weeks or days. Persistent delegation is designed to be long lived and makes role owners indistinguishable one from another, e.g., rights delegation to family of the car owner. It is obvious that the environment inside the car, e.g., mirror or seat position, does leak some information about the car occupant but addressing such issues is out of scope for us. We address privacy only from a protocol design perspective, i.e., the protocol run should not leak information about the role player in case of persistent users. Any other role can make ephemeral delegations of his access rights but persistent delegation can be done by the root only.

\textit{Rights revocation} can be done by the root or by the delegating user. Both persistent and ephemeral users can be revoked. Revocation requires a certificate revocation list that is maintained in the cloud so the car must have Internet connectivity, e.g., via 4G. While this is not a complicated demand for modern cars, it may be the case that in certain situations the car does not have such connectivity. In this situation, rights are to be revoked as soon as the car connects to the Internet or as soon as they expire based on the attributes.

\newcommand\barrow{\twoheadrightarrow}
\newcommand\sarrow{\rightarrow}

\newcommand\cert{\mathsf{Cert}}
\newcommand\man{\mathsf{Man}}
\newcommand\slr{\mathsf{Sel}}
\newcommand\sman{\mathit{mnf}}
\newcommand\sslr{\mathit{sel}}
\newcommand\oem{\mathsf{OEM}}
\newcommand\ecu{\mathsf{ECU}}
\newcommand\all{\mathsf{All}}

\newcommand\carid{\mathsf{CarID}}
\newcommand\pso{\mathsf{PsO}}
\newcommand\psu{\mathsf{PsU}}
\newcommand\pko{\mathsf{PkO}}

\newcommand\adr{\mathsf{adr}}
\newcommand\name{\mathsf{Name}}

\newcommand\rcvtag{\mathit{tag}_{\mathit{rcv}}}
\newcommand\rcvcount{\mathit{rcv}_{\mathit{count}}}
\newcommand\key{\mathsf{k}}
\newcommand\skey{\mathrm{K}_{\mathit{ses}}}
\newcommand\synckey{\mathrm{K}_{\mathit{sync}}}
\newcommand\smkey{\mathrm{K}_{m}}
\newcommand\mes{\mathrm{m}}
\newcommand\mesj{\mathrm{m_{j1939}}}
\newcommand\rnd{\mathit{rnd}}
\newcommand\sig{\mathit{Sig}}
\newcommand\enc{\mathit{enc}}
\newcommand\pbk{\mathit{pbk}}
\newcommand\mmod{\mathrm{mod}}
\newcommand\kd{\mathcal{KD}}
\newcommand\token{\mathit{atk}}

\newcommand\syncerr{\epsilon}
\newcommand\rdelay{\delta}

\newcommand\tecu{\mathit{t_{snd}}}
\newcommand\tecubit[1]{\mathit{t_{snd}^{#1}}}
\newcommand\tlocal{\mathit{t_{rcv}}}

\newcommand\erraut{\mathsf{AutFailed}}
\newcommand\retr{\mathsf{Retry}}
\newcommand\abort{\mathsf{RqAbort}}
\newcommand\rqkey{\mathsf{rqskey}}

\newcommand\msgid{\mathit{ID_m}}

\newcommand\car{\mathrm{Car}}
\newcommand\shcenter{\mathrm{ShCenter}}
\newcommand\sign{\mathsf{Sig}}
\newcommand\sigm{\mathsf{sig}}
\newcommand\ibs{\mathsf{IdSig}}
\newcommand\gs{\mathsf{GrSig}}
\newcommand\ttag{\mathsf{tag}}
\newcommand\pub{\mathit{Pb}}
\newcommand\priv{\mathit{Pv}}

\newcommand\ckey{\mathit{K}}
\newcommand\rights{\mathit{raccess}}
\newcommand\func{\mathit{func}}
\newcommand\ltime{\mathit{ltime}}

\newcommand\tstamp{\mathit{time}}

\newcommand\hash{\mathcal{H}}
\newcommand\mac{\mathsf{MAC}}
\newcommand\regdata{\mathrm{regdata}}
\newcommand\rfkey{\mathrm{RFKey}}
\newcommand\rand{\mathrm{rnd}}
\newcommand\cchain{\mathit{\mathcal{CP}\!\mathit{ath}}}

\newcommand\req{\mathrm{req}}

\newcommand\helper{\mathscr{H}}

\newcommand\info{\mathsf{rights}}
\newcommand\siginfo{\mathsf{sig\_rights}}
\newcommand\siginfocar{\mathsf{sig\_car\_rights}}
\newcommand\from{\mathrm{owner\!\!:}}
\newcommand\master{\mathrm{master}}

\newcommand\tstart{\mathit{T_{start}}}
\newcommand\tstop{\mathit{T_{stop}}}

\subsection{Cryptographic tools}

\textit{Symmetric primitives.} Our protocol makes use of standard Message Authentication Codes (MACs) and symmetric encryption which are instanced by SHA2 and AES in our proof-of-concept implementation. Besides these, we use more advanced cryptographic building blocks such as identity-based signatures and group signatures. We discuss these in more detail next. While symmetric primitives are present in all protocol actions along with asymmetric primitives, they play an exclusive role in the \textit{on-the-fly} execution procedure which is designed for fast interaction with the car.

\textit{Public-key primitives.} Besides the more complex identity-based and group-based primitives that we discuss next, our protocol uses regular \srevtext{public-key cryptographic functions. These are generally used in practice to establish a secure communication channel between participants by facilitating the exchange of a secret session key (which is later used for symmetric encryption).} To achieve this, we can rely on RSA \cite{Rivest78} encryption or, for a more compact representation, on the elliptical curve version of the Diffie-Hellman key-exchange \cite{Diffie76} \srevtext{, i.e., ECDH. Currently, 224-256 bit ECDH keys are viewed as the security equivalent of 2048-3072 bit RSA keys. While RSA leads to larger keys when compared to the elliptical-curve Diffie-Hellman, these are still easily manageable by modern smartphones.} \srevtext{We later show in the experimental section that the computational overhead induced by RSA is of little concern while manipulating keys of a few thousand bits is even less of a problem for modern smartphones that have several giga-bytes of RAM. }

\newcommand\iset{\mathsf{Setup}}
\newcommand\igen{\mathsf{KeyDer}}
\newcommand\isig{\mathsf{Sign}}
\newcommand\iver{\mathsf{Ver}}
\newcommand\imsk{\mathsf{msk}}
\newcommand\isk{\mathsf{sk}}
\newcommand\ipk{\mathsf{pk}}

\newcommand\ggen{\mathsf{Gen}}
\newcommand\gsig{\mathsf{Sign}}
\newcommand\gver{\mathsf{Ver}}
\newcommand\gtrace{\mathsf{Trace}}
\newcommand\ggmsk{\mathsf{gmsk}}
\newcommand\ggsk{\mathsf{gsk}}
\newcommand\ggpk{\mathsf{gpk}}

\textit{Identity-based signatures} (IBS) provide a more flexible framework which removes the need for exchanging digital certificates. The idea of identity-based signature originates from Shamir \cite{Shamir84}, for a more comprehensive introduction we refer the reader to \cite{Kiltz09}. 
In our proof-of-concept implementation, \srevtext{we consider both the original Shamir scheme as well as} the Guillou-Quisquater scheme \cite{Guillou90} which is part of the ISO/IEC 14888-2:2008 standard and is commonly proposed for use in embedded devices such as smart-cards \cite{Guillou01}. While this scheme is a bit more computational intensive than the regular RSA, it can be easily handled by modern Android devices (as we discuss in the experimental section) and it removes the need for digital certificates. \srevtext{To clarify the functionalities behind an identity-based signature, we provide next a generic description for it. For brevity, the concrete description of the two identity-based schemes is moved to Appendix A. An identity-based signature scheme consists of the following four algorithms:
\begin{enumerate}

\item $\iset(k)$ is the key setup algorithm which outputs the master secret key $\imsk$ and the global parameters $\ipk$. 

\item $\igen(\imsk, I)$ is the key generation algorithm which uses the master secret key $\imsk$ and the identity of the user $I$ to output the private key of the user. 

\item $\isig(\isk, m)$ is the signature generation algorithm which uses the secret key $\isk$ on the message $m$ to return the signature $\sigma$. 

\item $\iver(\ipk, I, m, \sigma)$ takes as input the system global parameters $\ipk$, the identity of the user $I$, the message $m$ and the signature $\sigma$ and returns true or false accordingly. 

\end{enumerate}
}

\textit{Group signatures} (GS) are used in order to provide the anonymity for group members. That is, the receiver of the signature can verify that it originates from a group member, but he cannot trace the particular group member. We use the scheme proposed by Boneh et al. in \cite{Boneh04}. Technical details are dense, we stick to a brief formalism that help us clarify the operations required by the group signature. The group signature is a collection of three algorithms:

\begin{enumerate}
\item $\ggen(n)$ is the key generation algorithm that takes the number of group users $n$ and returns the group secret master key $\ggmsk$, the group public key $\ggpk$ and the vector containing the group secret keys $\ggsk=\{\ggsk_1, \ggsk_2, ..., \ggsk_n\}$ that will be distributed to the group users,
\item $\gsig(\ggpk, \ggsk_i, m)$ is the signature algorithm that takes as input the group public key $\ggpk$, the secret key of the signer $\ggsk_i$ and a message $m$ then returns the signature $\sigma$,
\item $\gver(\ggpk, m, \sigma)$ is the verification algorithm which takes as input the group public-key $\ggpk$, the message $m$ and the signature $\sigma$ and returns \textit{true} if the signature is correct otherwise it returns $\perp$,
\item $\gtrace(\ggpk, \ggmsk, m, \sigma)$ is the tracing algorithm that can determine the signer based on the group secret key $\ggmsk$, the group public-key $\ggpk$, the message $m$ and the signature $\sigma$.
\end{enumerate}

Additionally, the group signature has mechanisms for revoking the keys. This will require updating the public key of the group. For simplicity, we skip formalism for this procedure.

As a general rule we assume that all signatures are time-stamped, i.e, they contain a timestamp and a loose time synchronization exists between all devices in the scheme. Such a requirement is in fact ubiquitous in Internet security and should not raise additional concerns for our scenario. To avoid overloading our notations, the timestamp is not explicitly mentioned in the signatures.

\newcommand\pk{\mathsf{pk}}
\newcommand\epk{\mathsf{pk}}
\newcommand\nonce{\mathsf{N}}
\newcommand\sesid{\mathsf{SID}}
\newcommand\sk{\mathsf{sk}}
\newcommand\gpk{\mathsf{gpk}}
\newcommand\gsk{\mathsf{gsk}}
\newcommand\own{\mathsf{Own}}
\newcommand\user{\mathsf{Usr}}
\newcommand\suser{\mathsf{usr}}
\newcommand\secown{\mathsf{Usr}}
\newcommand\ssecown{\mathsf{usr}}
\newcommand\driver{\mathsf{Drv}}
\newcommand\sdriver{\mathsf{drv}}
\newcommand\pdel{\mathsf{Del}}
\newcommand\psdel{\mathsf{PsD}}
\newcommand\psedel{\widetilde{\mathsf{PsDel}}}
\newcommand\spdel{\mathsf{del}}
\newcommand\edel{\widetilde{\mathsf{Del}}}
\newcommand\sedel{\widetilde{\mathsf{del}}}
\newcommand\role{\mathsf{Role}}
\newcommand\attrib{\mathsf{Atr}}
\newcommand\rig{\mathsf{Rights}}
\newcommand\sown{\mathit{own}}
\newcommand\scar{\mathit{car}}
\newcommand\aact{\mathsf{act\!\!:}}
\newcommand\execute{\mathsf{exec}}
\newcommand\conf{\mathsf{conf}}
\newcommand\inst{\mathsf{install}}
\newcommand\instsel{\mathsf{sel}}

\def\mwid{9cm}

\def\pwid{8.5cm}

\begin{figure}[ht!]
\footnotesize

\begin{minipage}{9cm}
\includegraphics[width=9 cm]{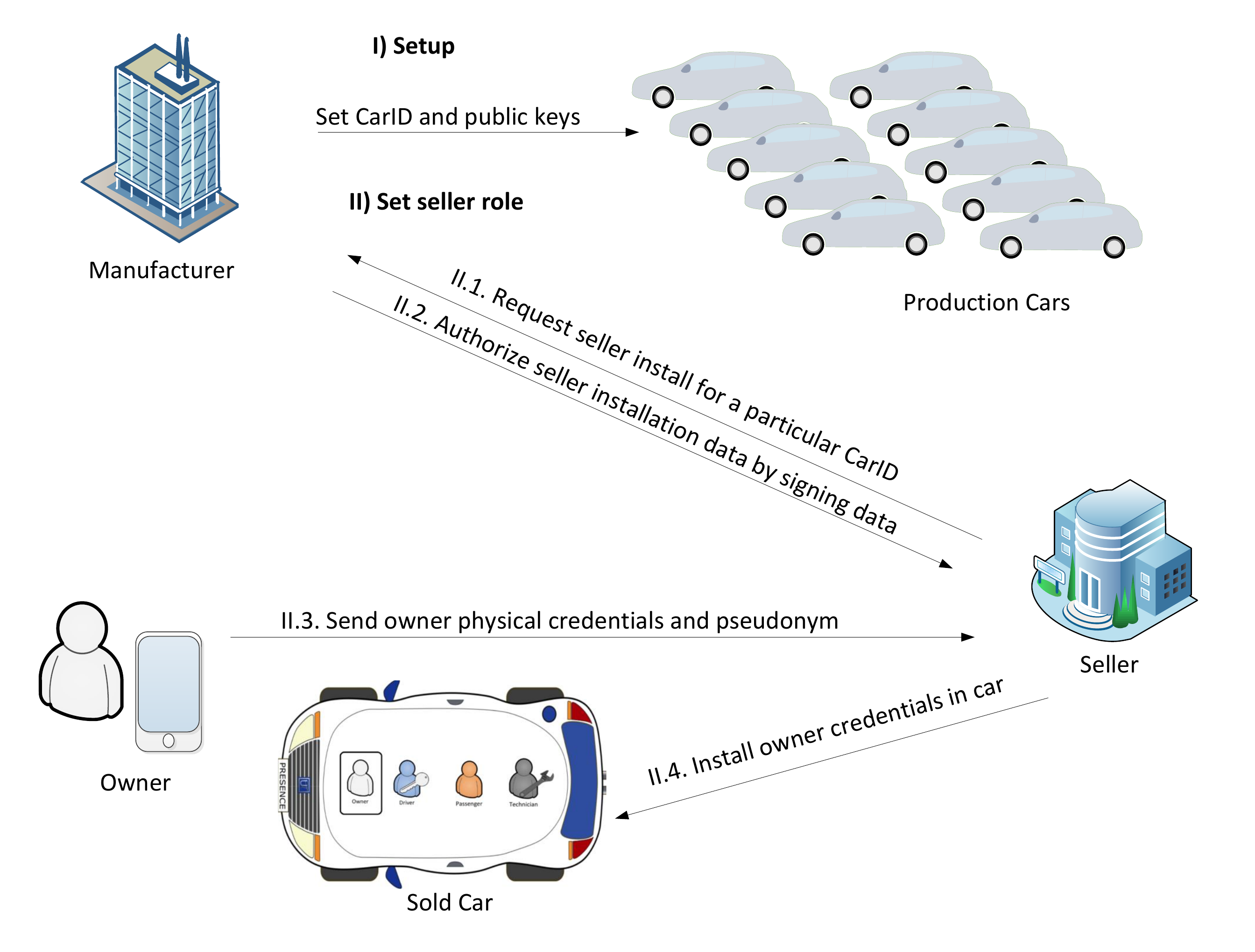}
\includegraphics[width=9 cm]{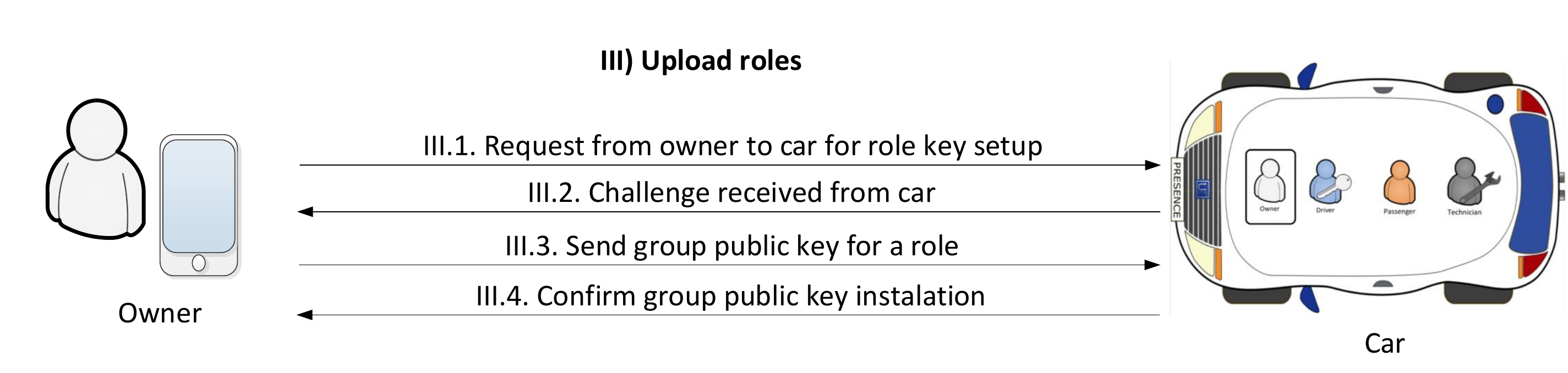}

\end{minipage}
\begin{minipage}{\pwid}
\renewcommand*{\arraystretch}{1.5}
\begin{tabular}{|p{\pwid}|}
		 \hline  
			\textbf{I) Setup (by manufacturer) } \\
			\\
			1. $\man \barrow \car$: $ \man, \carid,  \sk_{\scar}, \overline{\role} \times \attrib \times \overline{\rig} \} $\\
			\\		
			\hline
\end{tabular}
~~\\
~~\\
\begin{tabular}{|p{\pwid}|}
		 \hline  
			\textbf{II) Set root (by seller \& manufacturer) } \\
			\\
			1. $\slr \barrow \man$: $ m_{\sslr}=\{ \slr, \man, \carid \}, $	$s_{\sslr}=\ibs(\sk_{\sslr}, m_{\sslr}) $\\
			2. $\man \barrow \slr$: $ m_{\sman}=\{ \aact\instsel, \slr, \carid \}$,\\
			~~~~~~~~~~~~~~~~~~~~~$s_{\sman}=\ibs(\sk_{\sman}, m_{\sman}) $\\
			3. $\own \barrow \slr$: $ \mathit{owner~data}, m_{\sown} = \{ \pso, \tstart, \infty \} $\\
			4. $\slr \barrow \car$: $ m_{\sown}, s_{\sslr} = \ibs(\sk_{\sslr}, m_{\sown}), m_{\sman},  s_{\sman} $\\
			\\
			\hline
\end{tabular}
~~\\
~~\\
\begin{tabular}{|p{\pwid}|}
		 \hline 
			\textbf{III) Upload role public keys (by root only) } \\
			\\
			1. $\own \sarrow \car$: $ m'_{\sown}=\{ \nonce_{\sown}, \own, \carid\}$, \\
			~~~~~~~~~~~~~~~~~~~~~$s'_{\sown}=\ibs(\sk_{\sown}, m'_{\sown})$ \\
			2. $\car \sarrow \own$: $ m'_{\scar}=\{\nonce_{\sown}, \nonce_{\scar} \}, s'_{\scar}=\ibs(\sk_{\scar}, m'_{\scar}) $ \\
			3. $\own \sarrow \car$: $ m''_{\sown}=\{\nonce_{\sown}, \nonce_{\scar}, \role, \gpk, \tstart, \infty \},$ \\ 	~~~~~~~~~~~~~~~~~~~~$ s''_{\sown}=\ibs(\sk_{\sown}, m_{\sown}) $ \\
			4. $\car \sarrow \own$: $ m''_{\scar}=\{\aact\conf, m''_{\sown} \}, s''_{\scar}=\ibs(\sk_{\scar}, m''_{\scar}) $ \\
			\\
			\hline
\end{tabular}

\end{minipage}

\caption{Protocol procedures for car setup and group key upload}
\label{fig:setup_and_gpks}

\end{figure}

\begin{figure}[ht!]
\footnotesize
\centering
\includegraphics[width=9 cm]{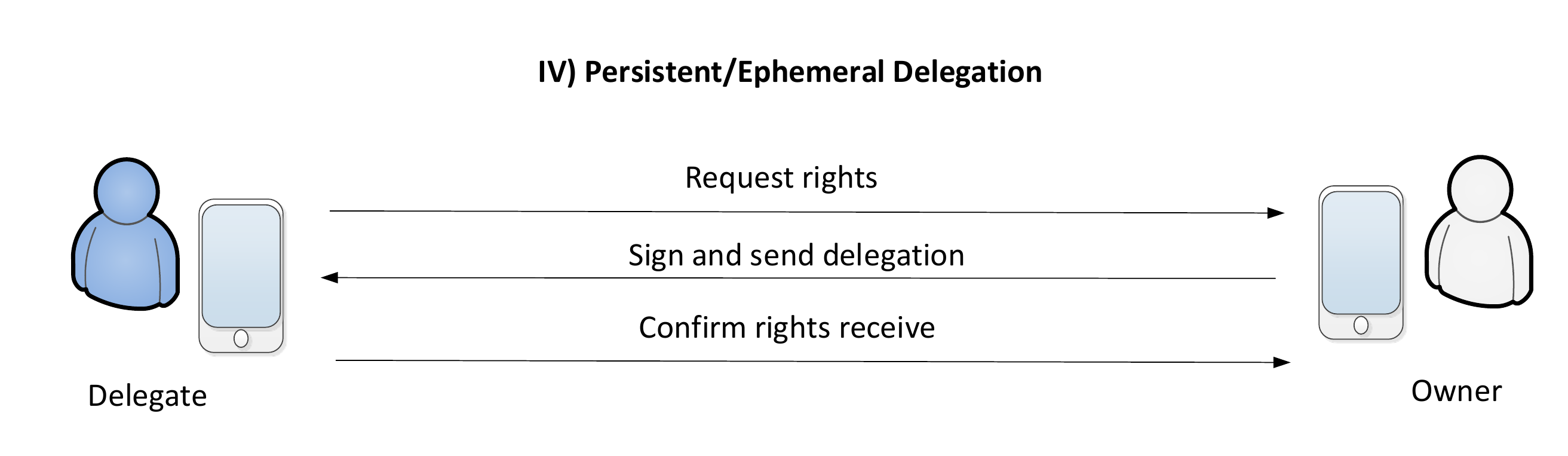}

\begin{minipage}{\mwid}
\renewcommand*{\arraystretch}{1.5}
\begin{tabular}{|p{\pwid}|}
		 \hline 
			\textbf{IV) Persistent or ephemeral delegation, i.e.,  $\user \in \{ \pdel, \edel \}$} \\
			\\
			1. $\user \sarrow \own$: $ m'_{\suser}=\{\psu,  \role, \attrib, \tstart, \tstop \}$,   \\
			~~~~~~~~~~~~~~~~~~~~~~~~~~~~~~~~$m''_{\suser}= \{ \epk_{\suser}, \nonce_{\suser}\}, s_{\suser}=\ibs(\sk_{\suser}, m'_{\suser}, m''_{\suser}) $ \\
			2. $\own \sarrow \user$: $ m'_{\sown}=\{\nonce_{\sown}, \nonce_{\suser}, \{ \token \}_{\skey} , \lfloor s_{\suser} \rfloor_{64} \}$,\\
			~~~~~~~~~~~~~~~~~~~~~~~~~~~~~~~$s'_{\sown}=\gs(\sk_{\sown}, m'_{\sown}) $ \\
			~~~~~~~~~~~~~~~~~~~~~~~~~~~~~~~$\token =\begin{cases} 
			\sk[\role]   \mathit{~if~} \user = \pdel \\
			s''_{\sown}=\gs(\sk_{\sown}, m'_{\suser}) \mathit{~if~} \user = \edel 
			\end{cases}$  \\
			3. $\user \sarrow \own$: $  m'''_{\suser}=\ibs(\sk_{\suser}, \{ m'_{\suser}, m'_{\sown} \} )$\\
			4. $\own \sarrow \user$: $  m''_{\sown} = \{ \skey \}_{\epk_{\suser}}$\\
			~~\\
			\\
			\hline
\end{tabular}
\end{minipage}

\caption{Protocol procedures for persistent and ephemeral rights delegation}
\label{fig:delegate}
\end{figure}

\subsection{Protocol steps}

The procedures required by the protocol are discussed next.

\textit{Car setup at the manufacturer.} The procedures for setting up the car and installing the root (which is the owner) are graphically depicted in Figure \ref{fig:setup_and_gpks}. We consider that these steps are done in a secure environment. Generally, this should be the case since if the production environment is insecure then the software on the ECUs may already be altered which may have more disastrous consequences. Still, if this is not the case, secure channels can be introduced during setup but this is out-of-scope for our work. We assume that during production, the manufacturer installs the secret key of the car, i.e., $\sk_{\scar}$, inside each car. Each car also has a unique identifier $\carid$ and the manufacturer is also responsible for installing the roles and rights which are expressed as a vector product $\overline{\role} \times \overline{\rig}$. Since we rely on identity-based signatures, all public keys will be derived by the car from the identities of the principals with which it interacts. Any other public parameters related to the identity-based cryptographic schemes will be installed in the car at this stage, i.e., step 1.

\textit{Setting the root owner during car purchase.}
The manufacturer is also responsible for giving rights to the seller as expressed in Protocol II from Figure \ref{fig:setup_and_gpks}. This happens by simply signing an installation message containing the identity of the seller and the identity of the car, i.e., $m_{\sman}=\{ \aact\instsel, \slr, \carid \}$. Both the request of the seller and the confirmation from the manufacturer are signed using an identity-based scheme, i.e.,  $s_{\sslr}=\ibs(\sk_{\sslr}, m_{\sslr}), s_{\sman}=\ibs(\sk_{\sman}, m_{\sman})$. 
When the car is purchased, the owner first presents the owner data (these are physical credentials as a passport or identification card, etc.). Due to legal issues it does not seem viable to hide driver's personal information from the seller, thus the owner has to present his physical credentials in some way. We assume that the seller is trustworthy and keeps the confidentiality of the new owner. The owner choses and presents a pseudonym $\pso$ and a public key $\pko$ and fixes the start of the contract as $\tstart$. The life-time of the purchasing contract is set to $\infty$ which seems a natural choice but can be changed according to practical needs. We use pseudonyms to assure driver's privacy in front of the car manufacturer. The seller $\slr$ verifies the legal information and if all the criteria are met, it installs the owner data inside the car by sending the owner request message $m_{\sown}$ with its signature $s_{\sslr}$ . For simplicity we also included here the messages for setting the identity of the seller inside the car, i.e., $m_{\sman},  s_{\sman}$, but these can be set as well at some previous stage. We assume that the seller installs the data received from the manufacturer in a secure manner inside the car  (the secure channel is suggested by the double arrow $ \barrow$). \revtext{If the setup needs to be done via an insecure port, such as OBD, we assume that this is done in a secure environment, e.g., an authorized garage. }

\textit{Setting group public keys.} This procedure is graphically suggested in Protocol III from Figure \ref{fig:setup_and_gpks}. The owner is the only entity entitled to add group public keys.
He starts a challenge-response interaction with the car by sending a message with a nonce $\nonce_{\sown}$, his identity $\own$ and the car $\carid$. This message is signed with an identity-based scheme as $s'_{\sown}=\ibs(\sk_{\sown}, m'_{\sown})$. The car replies with a message containing its own nonce $ \nonce_{\scar}$ and this message is as well signed as $s'_{\scar}=\ibs(\sk_{\scar}, m'_{\scar})$. The owner responds in $m''_{\sown}$ by including both the nonce that he sent and the one received from the car, the message also contains the name of the role $\role$ and the group public-key $\gpk$ along with start time $\tstart$ and validity period (we consider that validity is indefinite $\infty$ since the roles are persistent but this can be changed according to specific needs). The car confirms that the group public-key of the role was installed by signing message $m''_{\sown}$  and a confirmation tag $\aact\conf$.

\textit{Rights delegation scenario.} Both the persistent and ephemeral rights delegation procedures are suggested in Figure \ref{fig:delegate}, i.e., $\user \in \{ \pdel, \edel \}$.  The distinction between the two cases is in step 2 where the content of the authentication token $\token$ is different between the two. A persistent user will receive a group public key while ephemeral users will receive a signed proof of their execution rights. For the later case, the delegation holds for a specified amount of time between $\tstart$ and $\tstop$. We now describe the rest of the protocol which is identical regardless of the case. The user $\user$ makes a request to the owner by sending his pseudonym $\psu$, an ephemeral public-key $\epk_{\suser}$, a nonce $\nonce_{\suser}$ and the role $\role$ and attributes $\attrib$ for which he requests the rights. The message is signed by the user with an identity-based signature, i.e., $s_{\suser}=\ibs(\sk_{\suser}, m_{\suser}) $. The owner replies with a message containing a nonce $\nonce_{\sown}$, the encrypted authentication token $\{ \token \}_{\skey} $ and a truncated value from the signature of the previous message, i.e., $\lfloor s_{\suser} \rfloor_{64}$. The message also contains the signature of the owner which is a group signature for the group that he is part of, i.e., $s_{\sown}=\gs(\sk_{\sown}, m_{\sown})$. The reason for encrypting the authentication token and not yet disclosing the key is for the confirmation that the token was received in the next step. Otherwise, a delegated user may claim that he did not receive the token. In the next step the user confirms this with an identity-based signature on both previous messages, i.e., $\ibs(\sk_{\suser}, \{ m'_{\suser}, m'_{\sown}\})$. Now the owner discloses the session key $\skey$ in an encrypted manner such that only the user (which has the ephemeral public-key) can decrypt in the last message $\{ \skey \}_{\epk_{\suser}}$. The authentication token will be used in the next procedure for executing functionalities on the car. 
\revtext{The ephemeral public-key $\epk_{\suser}$ can be a regular RSA public-key or, if a more compact representation is desired, an elliptical curve Diffie-Hellman session key, i.e., the scalar multiplication of an elliptical curve point $aP$. For the later case, the owner will reply in the last message with the corresponding  key share, i.e., $bP$ and the session key $\skey$ is extracted from the common Diffie-Hellman key $abP$.} \srevtext{In the experimental section we show the differences in terms of computational costs, but both the Diffie-Hellman and RSA operations are in the order of dozen milliseconds which is affordable for modern Android devices.}

\begin{figure*}[ht!]
\footnotesize

\centering
\includegraphics[width=14 cm]{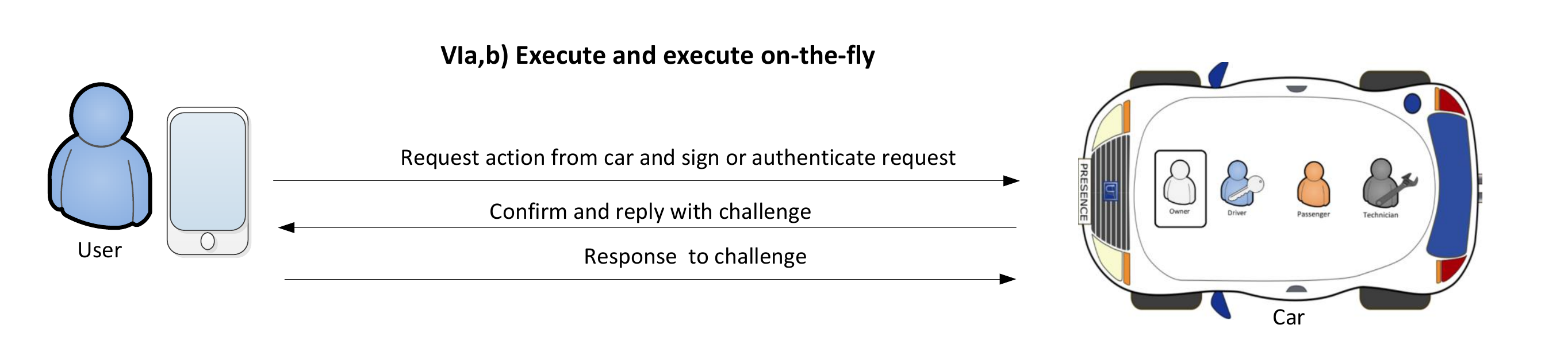}

\begin{minipage}{8.5 cm}
\renewcommand*{\arraystretch}{1.5}
\begin{tabular}{|p{8.25cm}|}
		 \hline 
			\textbf{V) Execute (by persistent or ephemeral role, i.e.,  $\user \in \{ \pdel, \edel \}$)}  \\
			\\

			1. $\user \sarrow \car$: $ m'_{\suser}= \{\nonce_{\suser}, \role, \attrib \},$ \\
			
			~~~~~~~~~~~~~~~~~~~~~~~$s'_{\suser}=\begin{cases} 
			\gs(\gsk[\suser], m_{\suser})  \mathit{~if~} \user = \pdel \\
			\ibs(\sk[\suser], m_{\suser}),  s''^{,iv}_{\sown}, m'^{,iv}_{\suser} \mathit{~if~} \user = \edel 
			\end{cases} 
			$\\
			
			2. $\car \sarrow \user$: $ m_{\scar}=\{\epk_{\scar}, \nonce_{\scar}, \sesid, \lfloor  s'_{\suser} \rfloor_{64} \}$, \\
			~~~~~~~~~~~~~~~~~~~~~~~~~~~~~~~~~~~~~~~~~~~~~~~~~~~~~~~~~~$s_{\scar}=\ibs(\sk_{\scar}, m_{\scar}) $ \\
			3. $\user \sarrow \car$: $ m''_{\suser}= \{ \{ \skey \}_{\epk_{\scar}}, \{ \aact\execute[i],  \lfloor  s_{\scar} \rfloor_{64} \}_{\skey} \}$, \\~~~~~~~~~~~~~~~~~~~~~~~~~~~~~~~~~~~~~~~~~~~~~~~~~~~~~~~~  $s''_{\suser} = \mac(\skey, m''_{\suser} )  $ \\		
			\\
			\hline
\end{tabular}
\end{minipage}~~~
\begin{minipage}{8.65 cm}
\renewcommand*{\arraystretch}{1.47}
\begin{tabular}{|p{8.15 cm}|}
		 \hline  
			\textbf{VI) Execute on-the-fly (by session key) } \\
			\\
			~~\\
			1. $\user \sarrow \car$: $ m_{\suser}=\{\sesid, \{ \aact\execute[i] \}_{\skey} \}, $\\
			~~~~~~~~~~~~~~~~~~~~~~~~~~~~~~~~~$s'_{\suser}=\mac(\skey, m_{\suser}) $ \\
			2. $\car \sarrow \user$: $ m_{\scar}=\{\nonce_{\scar}, \lfloor s'_{\suser} \rfloor_{64} \}, s_{\scar}=\mac(\skey, m_{\scar}) $ \\
			3. $\user \sarrow \car$: $ s''_{\suser}= \mac(\skey, s_{\scar} )  $ \\		
			~~\\
			~~\\
			\\			
			\hline
\end{tabular}
\end{minipage}

\caption{Protocol procedures for execution and on-the-fly execution}
\label{fig:act}
\end{figure*}

\textit{Execute functionality.} Procedures for triggering the execution of a functionality inside the car are graphically suggested in Figure \ref{fig:act}. The execute scenario calls for two distinct procedures that achieve the same goal. The first version relies on asymmetric primitives while the second (on-the-fly) relies on a session key and symmetric primitives alone. Obviously, the first procedure is more expensive and we assume that once a session key is established, only the second (faster) procedure is to be invoked. 
First, the user playing either a persistent or ephemeral user, i.e., $\user \in \{ \pdel, \edel \}$, sends a message $m'_{\suser}$ presenting his role and attributes along with a nonce $\nonce_{\suser}$ to assure freshness. This message comes along with signature $s'_{\suser}$ which is either a group signature (in case of persistent users)
 or an identity-based signature (in case of ephemeral users). For the later case, the user will also present the credentials based on which he acquired his rights from the owner, i.e., $s''^{,iv}_{\sown}, m'^{,iv}_{\suser}$ (these were received during the delegation procedure).
The car replies as challenge with an ephemeral public-key $\epk_{\scar}$, a nonce $\nonce_{\scar}$, a session identifier $\sesid$ included in $m_{\scar}$. These are signed by the car with an identity-based signature and presented as $s_{\scar}$. The message $m_{\scar}$ also includes a 64 bit truncation of the original signature from the user, i.e, $\lfloor  s_{\suser} \rfloor_{64}$.
Now the user generates a  session key $\skey$ which is encrypted with the ephemeral public-key $\epk_{\scar}$. The user presents the desired action on the car as an encrypted message, i.e., $\{ \aact\execute[i],  \lfloor  s_{\scar} \rfloor_{64} \}_{\skey}$ (note that this message also includes the last 64 bits of the car signature, i.e., $\lfloor  s_{\scar} \rfloor_{64}$). The message is accompanied by a regular MAC, i.e, $s''_{\suser}$, which is performed with the session key $\skey$. 
\revtext{
Similar to the previous protocol fragment, the ephemeral public-key $\epk_{\scar}$ can be a regular RSA public-key or a more compact Diffie-Hellman session key. In the later case $\epk_{\scar}$ is replaced by $aP$ and in the last message $\{ \skey \}_{\epk_{\scar}}$ is replaced by $bP$ from which the common session key $abP$ is extracted.
}

For the on-the-fly version of the protocol, the signatures are replaced with symmetric key Message Authentication Codes (MACs). The session identifier $\sesid$ links the on-the-fly execution with the session key $\skey$ from the previous procedure (the life-time of this session key can be hours, or days, depending on the practical circumstances). The user presents his request $\aact\execute[i]$ encrypted with the session key $\skey$ and authenticated with a MAC, i.e., $s'_{\suser}=\mac(\skey, m_{\suser})$. The car replies with a nonce $\nonce_{\scar}$ and this message along with the truncated value of $s'_{\suser}$ is also authenticated by a MAC in $s_{\scar}$. Finally, the user answers to this challenge with $s''_{\suser}$ which is a MAC computed on the previous message with the session key $\skey$.

We do not present additional procedures for \textit{rights revocation}. All of the included asymmetric primitives have well-known revocation mechanisms. This includes the group signature in \cite{Boneh04} for which the procedure is less obvious. While deploying such revocation mechanisms is not straight-forward, e.g., the car needs to keep a revocation list and update it accordingly, adding more details here is  out of scope for our work. \revtext{We also do not insist on other technicalities such as how to revoke owners or how to facilitate the resale of the car since adding protocol fragments is easy for any such action but will contribute little to the main concept from this work.}

\subsection{Adversary model and security arguments}

As adversary model, we consider the general Dolev-Yao \cite{Dolev83} intruder that has full control over the communication channel, i.e., he can eavesdrop, replay, inject or modify existing packets. Specific attacks related to the software implementation are out of scope for the current analysis. But as for future, more practical embodiments of our work, the use of specific Android security mechanisms or relying on hardware security, e.g., TPM 2.0 functions, may be projected.

Since we rely on existing cryptographic blocks that are assumed to be secure, rather than deriving more complicate cryptographic proofs, we consider that a proof by formal analysis (model-checking) offers better support for the security of our protocol. We choose to rely on the IF language for modelling which is the base language for the three model-checkers of the AVISPA platform \cite{Armando05}. In particular, we choose to rely on the CLAtse model-checker \cite{Turuani06} from the AVISPA platform. Model-checkers assume the underlying cryptographic blocks to be perfect and model the intruder as a Dolev-Yao adversary. For brevity, we choose to model the last 2 protocol fragments V) and VI), i.e., execute (by persistent or ephemeral role) and execute on-the-fly (by session key) since these are the actual protocol components that grant access to the car. Modelling the entire protocol would require a large amount of work and would be out-of-scope for the technological readiness level that we target in the current work, i.e., proof-of-concept. 
By using the IF language of the  AVISPA platform \cite{Armando05},  we model each protocol step as a transition from the left-hand side (LHS) facts to the right-hand side (RHS) facts. The LHS and RHS are conjunctions of positive and negative facts and are not persistent (the RHS suppresses the LHS). The Dolev-Yao adversary is modelled by the \emph{iknows} predicate which cumulates facts in a persistent manner, i.e., the intruder never forgets what he learns.

The first model that we analyze is the simple on-the-fly execution. We defined two actions, i.e., \emph{open car} and \emph{start car}, and ask the model checker if it can produce a trace that can trigger a \emph{start} of the car given that the honest user is set to \emph{open} the car. This covers the scenario in which the adversary can manipulate the commands of the genuine user. The model-checker answered that the protocol is safe. To test the correctness of our model we also added the session key $\skey$ to the intruder knowledge a case in which the model-checker immediately returned the attack (this proves that if the session key would have been leaked by any mean, the intruder would have been able to start the car). We also checked the consistency of the model by verifying that the genuine user can set the car in the \emph{open} open state which proved to be correct. 

Figure \ref{fig:usr_trace} shows the trace output by the model-checker when we tested that the genuine user can open the car (in case of the adversary attack, there is no trace since an attack cannot be found and the model is reported as safe). The output trace shows that the intruder $i$ mediates the communication channel by intercepting all messages. Compound messages are built with the \emph{pair} operator and symmetric encryption is modeled by the \emph{scrypt} predicate. Finally, the car reaches the state \emph{state\_car(2,usr,sid,kses,open,n4(NC))} which means that the user \emph{usr} managed to \emph{open} the car under session key \emph{kses} and a random challenge nonce \emph{n4(NC)} which is generated as a fresh symbolic term by the model checker. 

We then proceed to the analysis of the execute procedure by persistent or ephemeral roles. The AVISPA toolset does not offer specific support for identity-based signatures or group signatures but formally speaking they do not differ in terms of the signing/verification procedures from regular signatures (the differences are in how the keys are derived and linked to an identity). In the IF language, a signature is modelled as encryption with the inverse of the public-key (similar to the RSA mechanism). For example, the response of the car in step 2 of protocol v) is symbolically expressed as \emph{iknows(crypt(inv(PkCar), pair(PkCarE, pair(NC, pair(SID , crypt(inv(PkUsr), pair(NU, pair(Role, Atr))))))))}. Here, \emph{crypt(inv(PkCar),\_)} denotes the signature of the car on the message. We conducted similar tests as in the case of the on-the-fly procedure and the protocol proved to be safe.

\begin{figure}
\begin{center}

\scriptsize
\begin{Verbatim}[frame=single]
 
 i -> (car,4):  pair()
 (car,4) -> i:  pair(n4(NC),
            pair(scrypt(kses,pair(sid,open)),
            scrypt(kses,pair(n4(NC),
            scrypt(kses,pair(sid,open))))))
            & Remove state_car(0,usr,sid,kses,open,nc);
            & Add state_car(1,usr,sid,kses,open,n4(NC));
            & Built from trans2

 i -> (usr,3):  pair()
 (usr,3) -> i:  pair(sid,pair(scrypt(kses,open),
            scrypt(kses,pair(sid,scrypt(kses,open)))))
            & Remove state_usr(0,usr,sid,kses,open,nc);
            & Add state_usr(1,usr,sid,kses,open,nc);
            & Built from trans1

 i -> (usr,5):  pair(scrypt(kses,pair(n4(NC),
            scrypt(kses,pair(sid,open)))),
            pair(n4(NC),scrypt(kses,pair(sid,open))))
 (usr,5) -> i:  scrypt(kses,scrypt(kses,pair(n4(NC),
            scrypt(kses,pair(sid,open)))))
            & Remove state_usr(1,usr,sid,kses,open,nc);
            & Add state_usr(2,usr,sid,kses,open,n4(NC));
            & Built from trans3

 i -> (car,6):  scrypt(kses,scrypt(kses,pair(n4(NC),
            scrypt(kses,pair(sid,open)))))
 (car,6) -> i:  pair()
          & Remove state_car(1,usr,sid,kses,open,n4(NC));
          & Add state_car(2,usr,sid,kses,open,n4(NC));
          & Built from trans4
 
\end{Verbatim}

\end{center}
\caption{Output trace for regular user connection to the car}
\label{fig:usr_trace}
\end{figure}

\srevtext{Another type of attack at the protocol level that is worth considering is privilege escalation by which an attacker with certain privileges may try to perform an operation which he is not entitled to perform. This may include a manufacturer that tries to de-anonymize the users or a passenger that tries to achieve driver rights on the car. The only possibility to de-anonymize users is by using the traceability functionality of the signature designed by Boneh et al. \cite{Boneh04} which would require access to the group manager secret key. Our implementation delegates this capability to the car owner, emphasizing on the ownership rights (other implementations may delegate this to a trusted authority). Currently, the scheme of  Boneh et al. \cite{Boneh04} is considered secure, so it would be out-of-scope for this work to bring a new proof that an adversary may not perform this attack. The same remark is available for the suggested privilege escalation by a passenger as he will not be able to generate a signature for a role with higher privileges as long as the group signature scheme is secure.

There are of course many other side-channels from which the identity of drivers may be inferred. For example, driving patterns such as driving time or specific driving behaviors that can be recorded from on-device sensors can be used to infer the driver's identity. Recent research has proved that accelerometer data can be used for this purpose, e.g., \cite{Ly13}, \cite{Enev16}, \cite{Li18a}. It is however out-of-scope for the current research to address all these possible leakages as our work is focused on protocol design and implementation only. For a fully secure solution, one will need Android devices to offer resilience to such leakages.  
}

\section{Experiments}

In this section we discuss experimental results on Android phones and Infotainment units as well as on automotive-grade controllers. We discuss both computational requirements for some of the cryptographic primitives that we use as well as the protocol running time for several of the procedures that we previously described.

\subsection{Android and vehicle on-board unit implementation}

\begin{figure}[t!]
\centering
\begin{minipage}{9cm}
\centering
\includegraphics[width=9 cm]{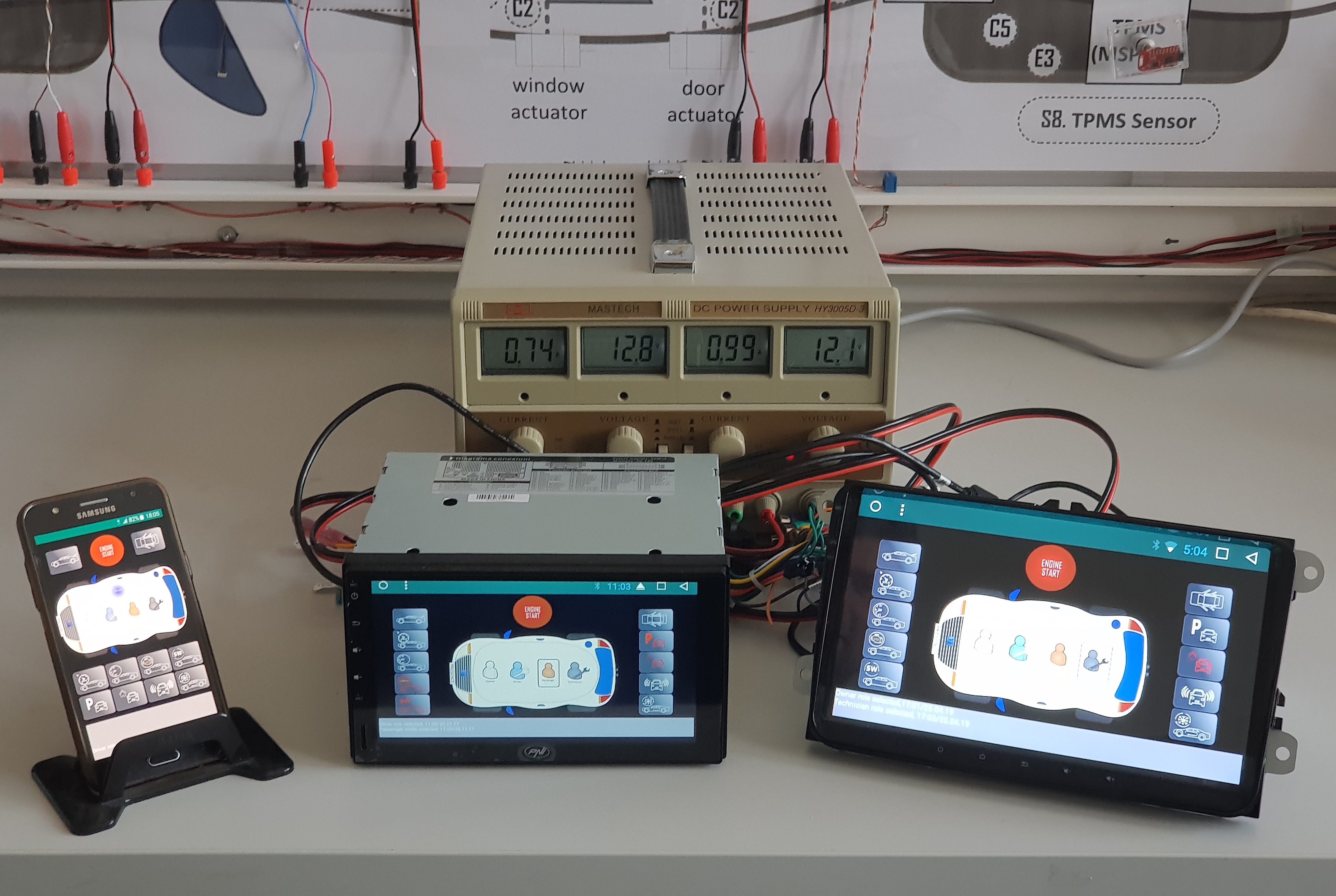}
\caption{Android headunits from our experiments}
\label{fig:headunit}
\end{minipage}
\end{figure}

Figure \ref{fig:headunit} depicts the experimental setup with the after-market Android headunits on which we deployed our implementation.

We have implemented in Android Studio the last three procedures of our protocol: persistent and ephemeral delegation, execute and execute on-the-fly. We tried to keep our implementation as simple and scalable as possible. Therefore, the protocol implementation relies on a simple finite state machine. The state machine consists of three states for each of the execute and execute on-the-fly procedure, and of four states for the delegation procedure, adhering to the previous formal protocol description. The states correspond to the messages that are exchanged during the procedures. For the group signature scheme we have used the Pairings\_in\_C library \footnote{https://github.com/IAIK/pairings\_in\_c} \cite{Unterluggauer14}. The library also contains an Android Demo that implements the group signature by Boneh et al. \cite{Boneh04}, which was easily adapted and integrated in our protocol. The identity-based signature scheme used in our protocol was independently implemented by us, while the rest of the cryptographic functions are from standard Java libraries. 

For the NFC communication, we have used the NFC card reader and NFC card emulation modes. As basis for our NFC implementation we used various samples from Android CardReader and Android CardEmulation Sample provided by Google \footnote{https://github.com/googlesamples/}. The payload size of the NFC frames that were transmitted between the device running in card reader mode and the device running in card emulation mode was 254 bytes. Hence, the messages exchanged in the implemented procedures were divided in several NFC frames.

Wi-Fi communication is based on TCP IP and we used two sockets, a server socket that listens the incoming connections requests and a client socket that initializes the connection. In our application, the smartphone is configured as a client and the head unit is configured as a server. The headunit also plays the role of the access point and mobile-phones connect directly to it.

For our experimental evaluation the on-board unit is represented by the TC297 microcontroller clocked at 300MHz and equipped with 8MB of flash and 728KB of RAM. We evaluated the computational performance of the TC297 by measuring the execution time for the basic building blocks of our protocol. We base our implementations on the Miracl (Multiprecision Integer and Rational Arithmetic Cryptographic Library) library \footnote{https://github.com/miracl/MIRACL}.

\subsection{Results}

The computational time of the required cryptographic primitives on several platforms is summarized in Table \ref{tab:comp_time}. The Shamir and GQ signature implementation was developed by us in C++ and Java for the Infineon controller and Android devices. The rest of the cryptographic functions come from the aforementioned libraries. 

On the Android devices, the computational time for the Group Signature (GS) is the highest along with the 2048-bit version of the GQ signature and tops between 250-500$ms$. \revtext{For the GQ signature, it may be that our implementation can be further optimized but the speed difference is clearly in favour of using Shamir IBS. For setting the security parameters of the Shamir signature, we followed the recommendations in \cite{Bellare07} which point to a 1024-bit modulus with a 160-bit hash functions, which we extend to a 2048-bit modulus with a 256-bit hash function that should be appropriate for current needs.} For the Tricore in-vehicle unit the execution time becomes unacceptable with the 2048-bit version of GQ \revtext{and since a 1024-bit modulus would lower the security level we find that using Shamir IBS is the only viable option. The cryptographic libraries that we use on Infineon had no platform-specific optimizations. The GS and the 2048-bit version of GQ have similar run-times on Android while there is a bigger computational gap between the two when executed on the Infineon controller. This suggests that the C++ code for the Infineon platform can be further optimized to obtain similar performances.} As the IBS and GS are used less often, the protocol should cope for a real-world car access scenario. We assume that the on-the-fly execution, which relies only on symmetric-key cryptography is the regular way to access the car while the identity/group-based execution is only triggered once to establish a session key. The RSA has a shorter runtime than the GS and GQ, but of course these traditional building blocks do not offer the advantages of group or identity-based signatures. \revtext{We also include results regarding ECDH, the time to generate the key-pair   $(a, aP)$ denoted as \textit{GenKP} and the time to generate the secret key $abP$ denoted as \textit{GenSK}, to serve as a comparison to RSA. For this purpose we use regular Android cryptographic libraries from Spongy Castle, while for the Infineon controller we consider results from our previous work in \cite{Popa19}. The runtime is in general comparable with that of the RSA, though not surprising RSA encryption is still the fastest. On the embedded controller the results for RSA were somewhat poorer and given the larger key size it becomes somewhat clear that ECDH would be more suitable in this case.}

\begin{figure}[t!]
\scriptsize
\centering
\begin{minipage}{4cm}
\centering
\includegraphics[width=4 cm]{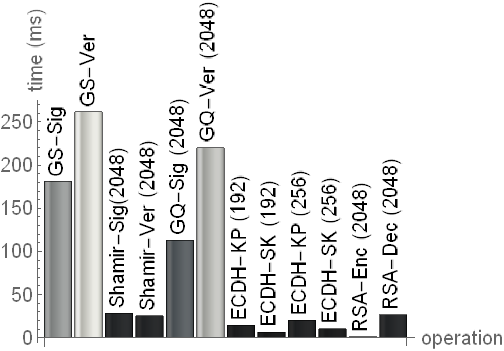}
(i)
\end{minipage}
\begin{minipage}{4cm}
\centering
\includegraphics[width=4 cm]{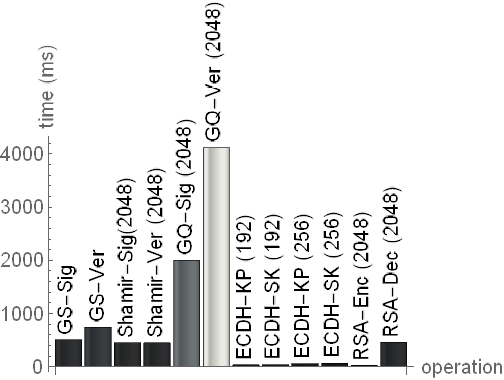}
(ii)
\end{minipage}

\begin{minipage}{4cm}
\centering
\includegraphics[width=4 cm]{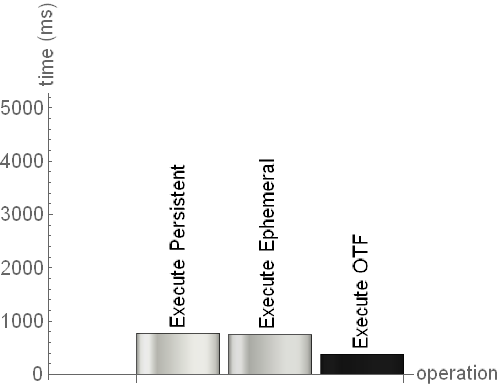}
(iii)
\end{minipage}
\begin{minipage}{4cm}
\centering
\includegraphics[width=4 cm]{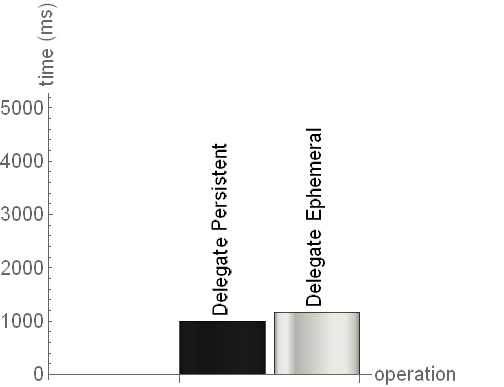}
(iv)
\end{minipage}

\caption{Execution time for: signatures on the ERISIN headunit (i) and Infineon Tricore in-vehicle board (ii), execute persistent/ephemeral (Shamir IBS) J5-Headunit 1(ERISIN) over WiFi (iii) and delegate (Shamir IBS) S5-S7 over NFC (iv)}

\label{fig:comp_time}
\end{figure}

\begin{table*}[h!]
\footnotesize
	\centering
    \begin{center}
	\caption{Computational time for cryptographic primitives on selected platforms (ms)}
	\label{tab:comp_time}
		 \begin{tabular}
		 {  >{\arraybackslash}p{2.65cm}|  r r r r r r r r r r r r r} 
		\hline \hline
		 \multirow{2}{*}{\backslashbox{Device}{Primitive}} &	  \multicolumn{2}{c}{\textbf{GS (254 bit)}} & \multicolumn{2}{c}{\textbf{Shamir (2048 bit)}} & \multicolumn{2}{c}{\textbf{GQ (2048 bit)}} 
		& \multicolumn{2}{c}{\textbf{ECDH (192 bit)}}   	& \multicolumn{2}{c}{\textbf{ECDH (256 bit)}} & \multicolumn{2}{c}{\textbf{RSA (2048 bit)}}  \\
		 &	 \textbf{Sign}& \textbf{Ver} & \textbf{Sign}  & \textbf{Ver}  
		& \textbf{Sign}  & \textbf{Ver} & \textbf{GenKP}  & \textbf{GenSK}
		& \textbf{GenKP}  & \textbf{GenSK} 		& \textbf{Enc}  & \textbf{Dec}

		 \\[0.5ex]	
		\hline  

\rowcolor[gray]{.95}

LG Optimus P700	& 259.21 	&364.35	&49.26	&46.49	&218.50	&424.80	&49.66 	&26.39 	&83.96  	&56.88  	&2.15  	&52.59 			\\[0.5ex]

Samsung S7 	&24.12  	&33.15  	&7.79  &8.02  &27.72  	& 57.18  			&15.15  	&13.81  	&46.37  	&15.17  	&0.24  	&8.80  	 \\[0.5ex]

\rowcolor[gray]{.95}
Samsung J5 SM-J500F &158.73  	&226.71  	&18.31  &16.21  &70.76  	&139.12 &13.58  	&5.95  &20.17  &10.39  	&0.60  	&17.07  	\\[0.5ex]

Samsung S5	& 68.09 &96.54 	&23.26 	&22.59 	&102.98 	&203.25 	&17.65 	&8.57 	&27.86 	&15.08 	&0.78 	&24.91   \\[0.5ex]

\rowcolor[gray]{.95} 
Headunit 1(Erisin)	&181.39  	&261.60  	&28.20  	&25.47  	&113.10  	&220.30  	&14.68  	&6.44  	&20.35  	&10.43  	&0.97  	&26.86  	 \\[0.5ex]

Headunit 2(PNI)	&176.33  	&254.08  	&27.35  	&24.70  &111.60  	&218.70  	&14.23 	&6.24  	&19.73  	&10.11  	&0.96  	&26.28  	\\[0.5ex]
\rowcolor[gray]{.95}
Tricore Tc297	&511.60  	&745.60  	&448.00  	&448.00	&2000.00	&4120.00 	&36.40  	&37.40  	&59.80  	&69.10  	&26.00  	&462.00  	 \\[0.5ex]
		
			\hline
			\hline
		\end{tabular}
	\end{center}
\end{table*}

\begin{table*}[h!]
\footnotesize
	\centering
    \begin{center}
	\caption{Computational/communication time for protocol procedures (ms)}
	\label{tab:prot_time}
		 \begin{tabular}
		 {  >{\arraybackslash}p{2.6cm} c r r r r r r r r r  } 
		\hline \hline
		\multirow{2}{*}		{Devices } &	\textbf{Com.} &  \multicolumn{2}{c}{\textbf{Execute Persistent}}& \multicolumn{2}{c}{\textbf{Execute Ephemeral}} & \textbf{Exec.}  & \multicolumn{2}{c}{\textbf{Delegate Persistent}} & \multicolumn{2}{c}{\textbf{Delegate Ephemeral}}  \\ 
		&	&\textbf{Shamir IBS} &\textbf{GQ IBS} &\textbf{Shamir IBS} &\textbf{GQ IBS} & \textbf{OTF} &\textbf{Shamir IBS} &\textbf{GQ IBS} &\textbf{Shamir IBS} &\textbf{GQ IBS}\\
		\hline   

\rowcolor[gray]{.95}
Phone2Phone(S5-S7) 			& NFC 	&\no  		&\no  		&\no  		&\no  		&\no  		&998.64 	&1256.42	&1165.64	&1423.42  \\[0.5ex]
S7-Headunit 1(Erisin) 	& WiFi	&758.48 	&892.55 	&923.87		& 1272.70	& 516.00	&\no 		&\no 		&\no 		&\no \\[0.5ex]
\rowcolor[gray]{.95}
S7-Headunit 2(PNI) 		& WiFi	&718.50 	&851.92		&1127.15 	& 1474.49 	 & 451.00	&\no 		&\no 		&\no 		&\no \\[0.5ex]
Phone2Phone (S5-S7)			& BT 	&366.83 	&567.42		&193.94 	& 523.42	 & 115.00	&\no 		&\no 		&\no 		&\no \\[0.5ex]
\rowcolor[gray]{.95}
J5-Headunit 1(Erisin) 	& WiFi	&771.85 	&1019.13 	&748.63	&1203.72 	&381.80 	&\no 		&\no 		&\no 		&\no \\[0.5ex]
J5-Headunit 2(PNI) 		& WiFi	&758.65 	&965.81 	&668.01 	&1121.61 	&274.50 	&\no 		&\no 		&\no 		&\no \\[0.5ex]

			\hline
			\hline
		\end{tabular}
	\end{center}
\end{table*}

In Table \ref{tab:prot_time} we summarize the complete run-time for several protocol procedures run between smartphones and head-units. These protocol fragments are tested over the three interfaces NFC, Bluetooth and WiFi. Since our car head-units did not support regular data transfer over Bluetooth, in this case we tested the execution only between two smartphones. However, the performance should be close to the case when a head-unit is used. Sharing rights is done over NFC due to increased security as it works on a shorter range and is harder to spoof. The request for execution to the car head-unit is done over WiFi. The execution runtime is around 1 second \revtext{(and generally less than 1 second with the more efficient Shamir IBS)}, which should be sufficiently fast, assuming that only the first execution is  done with the slower group or identity-based signatures. The rest of the executions are carried by the on-the-fly procedure taking only a few hundred milliseconds (since a common secret shared key exists). 
Figure \ref{fig:comp_time} summarizes in a graphic form on some of the computational times from Tables \ref{tab:comp_time} and \ref{tab:prot_time}.

\section{Conclusion}

The increased computational power of modern smartphones and their generous user-interface facilitates the implementation of various car access control functionalities and more exquisite protocols with advanced functionalities. These can benefit from state-of-the-art cryptographic building blocks such as identity-based cryptography or group signatures. While some of these require more computational power or build upon more expensive pairing-friendly elliptical curves, computational capabilities of modern smartphones and of high-end in-vehicle units are satisfactory for handling them. The provided experimental results prove that adoption is possible both on modern smartphones as well as on modern in-vehicle controllers, e.g., an Infineon TriCore car controller. At a minimum, the RBAC access control policy for car functionalities is within reach for most of the in-vehicle units on the market. With this research we hope to pave the way for addressing both security and privacy in car access control scenarios. \srevtext{Further improvements may consist in adding specialized hardware such as Trusted Platform Modules (TPM) or relying on  trusted execution environment such as ARM TrustZone that already exists on some mobile phones. Nonetheless, porting functionalities to wearable devices such as smart-watches or smart-glasses may also increase the usability of the solution. We leave these as potential directions for future works.}


\bibliographystyle{abbrv}
\bibliography{presto}

\section*{Appendix A - Description of the Shamir and Guillou-Quisquater Identity-based signature schemes}

\srevtext{
Since the native Android cryptographic libraries offer no support for the Shamir and Guillou-Quisquater identity-based signature schemes, we had to implement these separately (the source-code will be maintained on our project website). To clarify the algorithms we give their description in the syntax introduced in Section II.B.
}

The identity-based signature scheme proposed by Shamir \cite{Shamir84} consists in the following four algorithms:
\begin{enumerate}

\item $\iset(k)$ is the key setup algorithm which outputs the master secret key $\imsk$ and the global parameters $\ipk$. For this, it generates two random primes $p$, $q$ of $k$ bits in length, computes $n=pq, \phi(n)=(p-1)(q-1)$, sets integer $e \in Z_{\phi(n)}$ s.t. $\gcd(e,\phi(n))=1$  then computes $d=e^{-1} \mod \phi(n)$. The master secret key is $\imsk = \{n, d\}$ and the global public key is $\ipk = \{n, e, h \}$. Here $h$ stands for a hash function that maps the user name to an element of $Z_{\phi(n)}$, i.e., $h:\{0,1\}^{*} \rightarrow Z_{\phi(n)}$.

\item $\igen(\imsk, I)$ uses the master secret key $\imsk$ and the identity of the user $I$ to output the private key of the user. For this, it computes $h(I)^d \mod n$ and returns to each user the secret key $\isk = \{h(I)^d \mod n, n\}$ (the public key of each user is his identity, i.e., $I$). 

\item $\isig(\isk, m)$ is the signature generation algorithm which uses the secret key $\isk$ on the message $m$ to return the signature $\sigma$. For this, it selects random $r \in Z_n$, computes $t = r^e \mod n$, then the hash of $t$ concatenated with message $m$, i.e., $h=\mathit{hash}(t||m)$, then $s=h(I)^dr^h \mod n$. The signature is $\sigma = \{s, t\}$.

\item $\iver(\ipk, I, m, \sigma)$ takes as input the system global parameters $\ipk$, the identity of the user $I$, the message $m$ and the signature $\sigma$. To verify that the signature is correct the algorithm computes $s^e$ then checks if this is equal to $h(I)t^h \mod n$ and returns \emph{true} if so or $\perp$ otherwise.

\end{enumerate}

The Guillou-Quisquater \cite{Guillou90} identity-based signature scheme is a collection of four algorithms:

\begin{enumerate}

\item $\iset(k)$ is the key setup algorithm that generates the master secret key $\imsk$ and the public key $\ipk$. In case of the GQ algorithm, the $\iset$ algorithm, generates two random primes $p$, $q$, each having $k$ bits in length, it selects random integer $v \in Z_n, n=pq$, computes $\phi(n)=(p-1)(q-1)$ and $v^{-1} \mod \phi(n)$. The master secret key is $\imsk = \{n, v^{-1} \mod \phi(n)\}$ and the public key is $\ipk = \{n, v \}$.

\item $\igen(\imsk, I)$ is the key derivation algorithm that
uses the master secret key $\imsk$ and the identity of the user $I$ to generate his private key. In case of the GQ algorithm, the identity $I$ of a principal is mapped (by a publicly known redundancy function) to a number $J \in Z_n$ then the algorithm computes $B=J^{-v^{-1}} \mod n$. The user secret key is $\isk = \{B, J, v, n\}$ (since this is an identity-based scheme, the public key to verify the signatures of this user is $\ipk$ and the identity of the user $I$). 

\item $\isig(\isk, m)$ is the signature algorithm that takes as input the user's secret key $\isk$ and a message $m$ then returns the signature $\sigma$. The GQ signing algorithm selects a random $r \in Z_n$, computes $T = r^v mod n$, the hash of message $m$ denoted as $h$, then $d=J^{h}T^{v^l} \mod n$ and $t=r B^d \mod n$ where $l$ is an integer such that $v^l < m < v^{l+1}$. The signature is $\sigma = \{d, t\}$.

\item $\iver(\ipk, m, \sigma)$ is the verification algorithm which takes as input the public-key $\ipk$, the message $m$ and the signature $\sigma$ and returns \emph{true} if the signature is correct otherwise it returns $\perp$. To verify that the signature is correct, the algorithm derives $J$ from the identity $I$, computes $T' = J^d T^v \mod n$, computes the hash $h$ of messages $m$ and $d'=J^hT'^{v^l} \mod n$, then the verifier $d'' = J^{h+dv^l}t^{v^{l+1}} \mod n$ and checks if $d'=d''$ then returns \emph{true} if so or $\perp$ otherwise.

\end{enumerate}

\newpage

\begin{IEEEbiography}[{\includegraphics[width=1in,height=1.25in,clip,keepaspectratio]{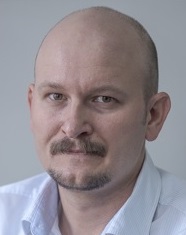}}]{Bogdan Groza} is Professor at Politehnica University of Timisoara (UPT). He received his Dipl.Ing. and Ph.D. degree from UPT in 2004 and 2008 respectively. In 2016 he successfully defended his habilitation thesis having as core subject the design of cryptographic security for automotive embedded devices and networks. He has been actively involved inside UPT with the development of laboratories by Continental Automotive and Vector Informatik. Besides regular participation in national and international research projects in information security, he lead the CSEAMAN project (2015-2017) and currently leads the PRESENCE project (2018-2019), two research programs dedicated to automotive security funded by the Romanian National Authority for Scientific Research and Innovation. 
\end{IEEEbiography}

\begin{IEEEbiography}[{\includegraphics[width=1in,height=1.25in,clip,keepaspectratio]{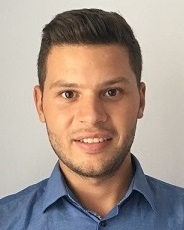}}]{Tudor Andreica} is a Ph.D. student at Politehnica University of Timisoara. He graduated his B.Sc and M.Sc studies in 2016 and 2018 respectively, from Polithenica University of Timisoara. Since 2015 he is working as software engineer at HELLA Romania focusing on the security of various in-vehicle systems. He was a research student in the CSEAMAN project and currently joined the PRESENCE project as a PhD student. His research interests are in the field of automotive cybersecurity. 
\end{IEEEbiography}

\begin{IEEEbiography}[{\includegraphics[width=1in,height=1.25in,clip,keepaspectratio]{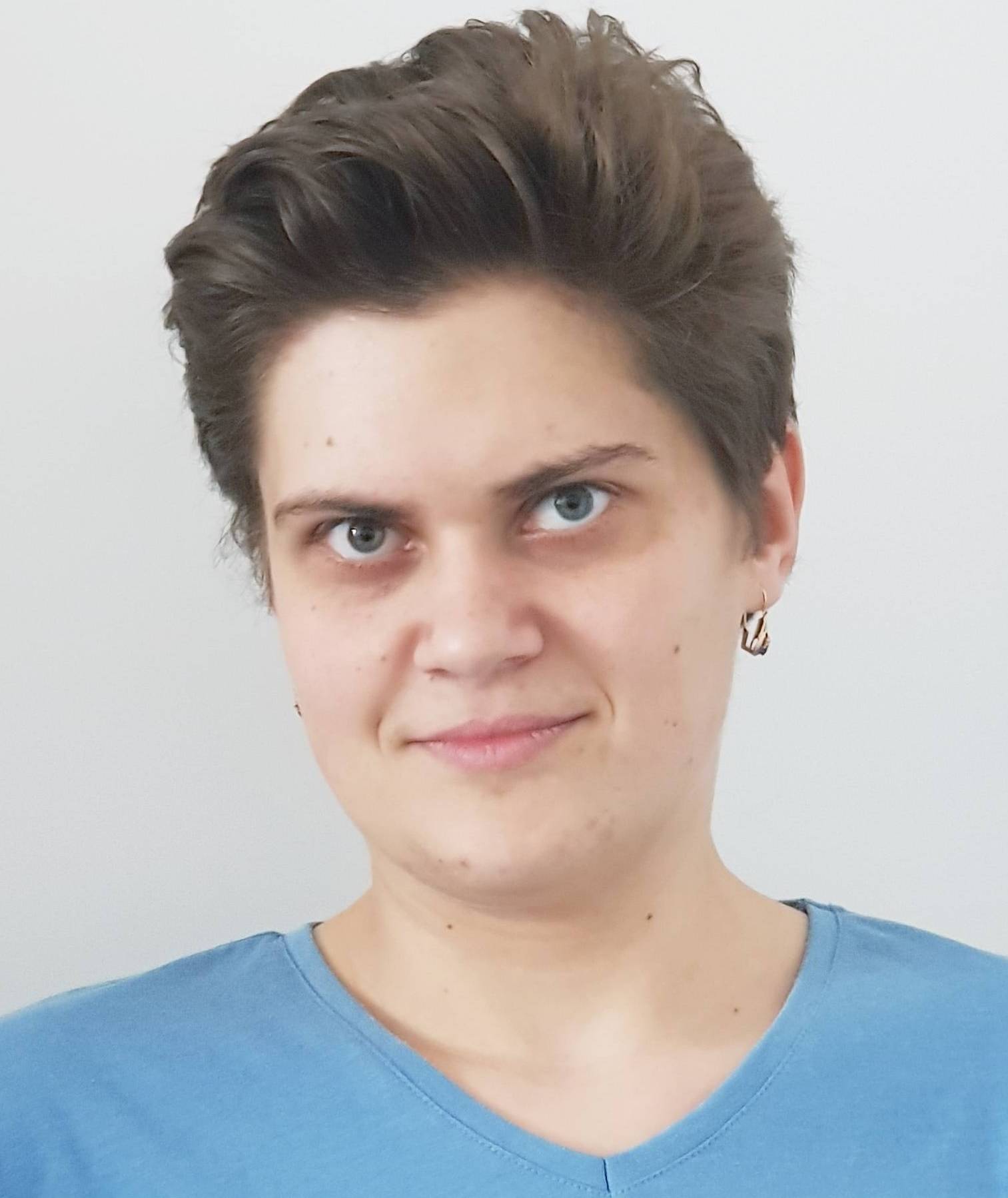}}]{Adriana Berdich} received the Dipl.Ing. and M.Sc. degrees from the Politehnica University of Timisoara (UPT), in 2017 and 2019, respectively, where she is currently pursuing the Ph.D. degree. From 2015 to 2018, she was a Software Developer in the automotive industry for Continental Corporation in Timisoara. She is also a research student in the Presence project focusing on environment-based device association and also continues as a software developer in the automotive industry for Vitesco Technologies focusing on power-train applications.
\end{IEEEbiography}

\begin{IEEEbiography}[{\includegraphics[width=1in,height=1.25in,clip,keepaspectratio]{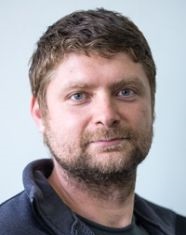}}]{Pal-Stefan Murvay} is a Lecturer at Politehnica University of Timisoara (UPT). He graduated his B.Sc and M.Sc studies in 2008 and 2010 respectively and received his Ph.D. degree in 2014, all  from  UPT. He has a 10-year background as a software developer in the automotive industry. He worked as a postdoctoral researcher in the CSEAMAN project and is currently a senior researcher in the PRESENCE project. He also leads the SEVEN project related to automotive and industrial systems security. His current research interests are in the area of automotive security. 
\end{IEEEbiography} 

\begin{IEEEbiography}[{\includegraphics[width=1in,height=1.25in,clip,keepaspectratio]{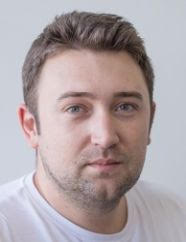}}]{Eugen Horatiu Gurban}  is Lecturer at Politehnica University of Timisoara (UPT). He received the Dipl.Ing. degree in 2007, MsC. degree in Automotive Embedded Software in 2009, and Ph.D. degree in 2014, all from UPT. He worked for 3 years (2007 - 2010) in the automotive industry as a software verification engineer and team leader at the R\&D Software Engineering Department of Autoliv Romania. He currently works as postdoctoral researcher in the PRESENCE project. 
\end{IEEEbiography}

\EOD

\end{document}